\documentclass[reprint,aps,dvipdfmx,amsmath,prb,superscriptaddress]{revtex4-1}

\usepackage{graphicx,color}
\usepackage{times}

\usepackage[breaklinks]{hyperref}
\hypersetup{setpagesize=false, colorlinks=true, citecolor=blue, urlcolor=blue, linkcolor=blue}

\renewcommand{\theequation}{\arabic{section}.\arabic{equation}}
\newcommand{\sinc}{\mathrm{sinc}}

\newcommand{\vac}{\mathrm{vac}}

\newcommand{\cpdc}{\mathrm{tri}}
\newcommand{\en}{\mathrm{e}}
\newcommand{\p}{\mathrm{p}}

\newcommand{\GSB}{\mathrm{GSB}}
\newcommand{\SE}{\mathrm{SE}}
\newcommand{\ESA}{\mathrm{ESA}}
\newcommand{\DQC}{\mathrm{DQC}}
\newcommand{\tr}{\mathrm{tr}}
\newcommand{\E}{\hat{E}}
\newcommand{\C}{C}
\newcommand{\dt}{\Delta t}
\newcommand{\D}{D}
\newcommand{\cm}{{}\,\mathrm{cm}^{-1}}

\newcommand{\fs}{{}\,\mathrm{fs}}

\newcommand{\nr}{\mathrm{(nr)}}
\renewcommand{\r}{\mathrm{(r)}}

\begin{document}
\title{Achieving two-dimensional optical spectroscopy with temporal and spectral resolution
using quantum entangled three photons}

\author{Yuta Fujihashi}
\altaffiliation{Present address: Department of Molecular Engineering, Graduate School of Engineering, Kyoto University, Kyoto 615-8510, Japan and PRESTO, Japan Science and Technology Agency, Kawaguchi 332-0012, Japan}
\thanks{fujihashi@moleng.kyoto-u.ac.jp}
\affiliation{Institute for Molecular Science, National Institutes of Natural Sciences, Okazaki 444-8585, Japan}
\affiliation{PRESTO, Japan Science and Technology Agency, Kawaguchi 332-0012, Japan}

\author{Akihito Ishizaki}
\thanks{ishizaki@ims.ac.jp}
\affiliation{Institute for Molecular Science, National Institutes of Natural Sciences, Okazaki 444-8585, Japan}
\affiliation{School of Physical Sciences, Graduate University for Advanced Studies, Okazaki 444-8585, Japan}

\begin{abstract}
Recent advances in techniques for generating quantum light have stimulated research on novel spectroscopic measurements using quantum entangled photons. One such spectroscopy technique utilizes non-classical correlations among entangled photons to enable measurements with enhanced sensitivity and selectivity. 
Here, we investigate spectroscopic measurement utilizing entangled three photons.
In this measurement, time-resolved entangled photon spectroscopy with monochromatic pumping [J.~Chem.~Phys. {\bf 153}, 051102 (2020).] is integrated with the frequency-dispersed two-photon counting technique, which suppresses undesired accidental photon counts in the detector and thus allows one to separate the weak desired signal.
This time-resolved frequency-dispersed two-photon counting signal, which is a function of two frequencies, is shown to provide the same information as that of coherent two-dimensional optical spectra. The spectral distribution of the phase-matching function works as a frequency filter to selectively resolve a specific region of the two-dimensional spectra, whereas the excited-state dynamics under investigation are temporally resolved in the time region longer than the entanglement time. 
The signal is not subject to Fourier limitations on the joint temporal and spectral resolution, and therefore, it is expected to be useful for investigating complex molecular systems in which multiple electronic states are present within a narrow energy range. 
\end{abstract}

\maketitle

\section{Introduction}

Quantum entanglement is one of the properties that is unique to quantum mechanics. When the state of the entire system cannot be described as a product of the quantum states of its constituent particles, such a system is referred to as being entangled. \cite{Schrodinger:1935kq}
The most common types of entanglement are the polarization entanglement of photon pairs and the spin entanglement of electron pairs. They also include correlations related to continuous quantities such as the position-momentum of two particles,\cite{Ou:1992re,Eisert:2003in,Braunstein:2005qu,Weedbrook:2012ga,Asavanant:2019ge} which was first discussed in the Einstein-Podolsky-Rosen paradox.\cite{Einstein:1935jx} Energy and charge transports in photosynthetic proteins were also discussed from the perspective of quantum entanglement. \cite{Thorwart:2009en,Sarovar:2010hs,Caruso:2010fx,Ishizaki:2010ft,Fassioli:2010kb,Whaley:2011qu,Ishizaki:2012kf}

Entangled states also play essential roles in state-of-the-art quantum technologies. \cite{Takeuchi:2014re,Walmsley:2015cn,Simon2016book,Pirandola:2018ad,Moreau:2019im} In the past few decades, advances in techniques for generating broadband frequency-entangled photons and shaping the time-frequency structures of entangled photons have stimulated research on novel spectroscopic measurements using entangled photon pairs. \cite{Georgiades:1995dd,Scarcelli:2003re,Yabushita:2004hy,Dayan:2004kg,Lee:2006id,Kalachev:2013kh,Upton:2013is,Kalashnikov:2016cl,Varnavski:2017eq,Villabona:2017en,Villabona:2020me,Paterova:2018me,Lee:2020mo,Szoke:2020fc} One such entangled photon spectroscopy technique utilizes non-classical photon correlations to enable measurements with enhanced sensitivity and selectivity when compared to conventional techniques based on classical physics. For instance, two-photon absorption induced by entangled photon pairs varies linearly rather than quadratically with light intensity. \cite{Gea1989:tw,Javanainen1990:li,Georgiades:1995dd,Dayan:2004kg,Kang:2020ef} It has also been argued that two-photon excitation in molecules can be manipulated to specific electronic states. \cite{Fei:1997es,Saleh:1998vl,Oka:2010if,Oka:2011dy,Schlawin:2012ba,Schlawin2013two,Schlawin:2013dq,Raymer:2013kj,Dorfman:2014jm,Munkhbaatar:2017se,Schlawin:2018ci,Oka:2018en,deJLeonMontiel:2019jt,Oka:2020en,Bittner:2020gr} Two-photon coincidence detection \cite{Hong:1987gm,Pittman:1995un} and double-crystal interference experiments \cite{Zou:1991in,Lemos:2014qu} have also been studied with respect to spectroscopic applications. 
In a typical coincidence scheme, one pair of entangled photons is employed as a probe field that is transmitted through the molecular sample. The remaining one is detected in coincidence. This type of measurement improves the signal-to-noise ratio. \cite{Scarcelli:2003re,Yabushita:2004hy,Kalachev:2013kh,Okamoto:2020lo} It is also possible to conduct infrared spectroscopy using visible detectors by exploiting the non-classical correlations between entangled photon pairs. \cite{Kalashnikov:2016cl,Paterova:2018me}

To date, experimental explorations have been limited to steady-state spectroscopic measurements as stated above. Given the growing need to understand dynamical processes in complex molecular systems and materials, it is important to extend entangled photon spectroscopy to time-resolved measurements. Pump-probe and stimulated Raman spectroscopic measurements with two-photon counting were theoretically proposed through a combination of biphoton spectroscopy with additional laser pulses. \cite{Dorfman:2014bn,Schlawin:2016er} In a previous study, \cite{Ishizaki:2020jl} we theoretically investigated the frequency-dispersed transmission measurement of an entangled photon pair that was generated using a monochromatic laser. It was demonstrated that the non-classical correlation between this photon pair enabled time-resolved spectroscopy using monochromatic pumping. However, transmission measurements are not background-free; weak nonlinear signals must be separated from the probe field that is transmitted through a sample. Therefore, the signal-to-noise ratio is limited by shot noise. Furthermore, it becomes difficult to detect nonlinear optical signals induced by photon pairs in regimes with low photon fluxes.

In this study, we investigate a spectroscopic method to overcome the difficulties associated with implementing time-resolved entangled photon spectroscopy. The central idea is to use entangled three photons \cite{Greenberger:1990be,Keller:1998fn,Wen:2007tr,Wen:2009im,Hubel:2010kp,Shalm:2013ia,Hamel:2014di,Agne:2017ob,Corona:2011ex,Krapick:2016ch,Moebius:2016ef,Zhang:2018ev,Cho:2018cf,Okoth:2019se,Dominguez:2020th,Ye:2020dz} and frequency-dispersed two-photon coincidence counting measurements. In this scheme, two of the three photons are irradiated into the molecular sample to induce a nonlinear optical process, while the remaining photon is detected in coincidence with the probe field transmitted through the sample. Coincidence-based transmission measurements suppress undesired accidental photon counts in the detector which measures the probe field. \cite{Scarcelli:2003re,Yabushita:2004hy,Kalachev:2013kh} 
Thus, this technique enables us to separate the genuine spectroscopic signal.
We show how the non-classical correlation among the entangled three photons can be exploited such that two-photon coincidence measurements can provide information on dynamical processes in molecules, similar to transmission measurements of an entangled photon pair. \cite{Ishizaki:2020jl}

This paper is organized as follows: In Sec.~II, we address the quantum states of the entangled three photons generated via cascaded PDC. \cite{Hubel:2010kp,Shalm:2013ia,Hamel:2014di,Agne:2017ob} We also describe the frequency-dispersed two-photon coincidence counting signal in the three photon state. In Sec.~III, we present numerical results to clarify the influence of entanglement times on the spectroscopic signals.
Section IV is devoted to the concluding remarks.

\section{Theory}
\subsection{Generation of entangled three photons via cascaded PDC}

One of the most widespread techniques for generating these quantum resources is parametric down-conversion (PDC). \cite{mandel1995optical} In this process, a photon originating from an input laser is converted into an entangled photon pair in a way that satisfies the energy and momentum conservation laws. In this work, we address entangled three photons generated through the cascaded PDC process with two nonlinear crystals, \cite{Hubel:2010kp,Shalm:2013ia,Hamel:2014di,Agne:2017ob} as shown in Fig.~\ref{fig:1}. 
In the primary PDC, the pump photon, which has a frequency of $\omega_\p$, passes through the first crystal and is split into a pair of daughter photons (photons~0 and 1) with frequencies of $\omega_0$ and $\omega_1$. In the second crystal, photon~0 serves as the pump field for the secondary conversion, creating a pair of granddaughter photons (photons~2 and 3) with frequencies of $\omega_2$ and $\omega_3$. 
It is noted that the conversion efficiency in the cascaded PDC is typically very low. 
For instance, in cascaded PDC with a combination of periodically-poled lithium niobate (PPLN) and periodically-poled potassium titanyl phosphate, the detected rate of entangled three photons is approximately 7 counts per hour. \cite{Shalm:2013ia}
At this photon count rate, an unrealistically long signal collection time may be required to measure the frequencies of photon for extracting the spectroscopic information.
However, the main purpose of this work is to give how the non-classical correlations among entangled three photons can be applied to time-resolved spectroscopy. 
A discussion on the experimental feasibility of the frequency-dispersed two-photon coincidence counting measurement using the cascaded PDC is beyond the scope of this paper.

For simplicity, we consider the electric fields inside the one-dimensional nonlinear crystals. In the weak down-conversion regime, the state vector of the generated three photons is written as \cite{Shalm:2013ia,Zhang:2018ev,Ye:2020dz}
\begin{align}
	\lvert \psi_\cpdc \rangle 
	\simeq
	\iiint d^3\omega
	f(\omega_1, \omega_2, \omega_3 ) 
	\hat{a}_1^\dagger(\omega_1) 
	\hat{a}_2^\dagger(\omega_2) 
	\hat{a}_3^\dagger(\omega_3) 
	\lvert \vac \rangle.
	\label{eq:state-vector}
\end{align}
The derivation is given in Appendix~A.
In the equation, $\hat{a}_{\sigma}^\dagger(\omega)$ denotes the creation operator of a photon of frequency $\omega$ against the vacuum state $\lvert \vac \rangle$. The operator satisfies the commutation relation $[\hat{a}_{\sigma}(\omega), \hat{a}_{\sigma'}^\dagger (\omega')]=\delta_{\sigma  \sigma'}\delta(\omega - \omega')$.
The three-photon amplitude, $f(\omega_1, \omega_2, \omega_3 )$, is expressed as 
\begin{align}
	f(\omega_1, \omega_2, \omega_3 ) 
	=
	\eta A_\p(\omega_1 + \omega_2 + \omega_3)
	\phi(\omega_1, \omega_2, \omega_3),
	\label{eq:three-photon-amplitude}
\end{align}
where $A_\p (\omega)$ is the normalized pump envelope and $\phi(\omega_1,\omega_2,\omega_3)= \sinc[\Delta k_1 (\omega_2 + \omega_3,\omega_1)L_1 /2]$
$\times \sinc[\Delta k_2 (\omega_2,\omega_3)L_2 /2]$ denotes the phase-matching function of the overall cascaded PDC process.
The momentum mismatch between the input and output photons in the $n$-th nonlinear crystal is expressed by $\Delta k_1(\omega, \omega')= k_\p(\omega+\omega') - k_0(\omega) -k_1(\omega')$ and $\Delta k_2(\omega, \omega')= k_0(\omega+\omega') - k_2(\omega) -k_3(\omega')$. The length of the $n$-th crystal is given by $L_n$. The momentum mismatches may be linearly approximated around the central frequencies of the generated beams, $\bar\omega_{\sigma}$, as in \cite{Zhang:2018ev,Ye:2020dz}
\begin{align}
	\Delta k_1(\omega_0, \omega_1) L_1
	&= 
	(\omega_0 - \bar\omega_{0} ) T_{\p0} + (\omega_1 - \bar\omega_1) T_{\p1}, 
	\label{eq:momentum-mismatch1}
\\[12pt]
	\Delta k_2(\omega_2, \omega_3) L_2
	&= 
	(\omega_2 - \bar\omega_2) T_{02} + (\omega_3 - \bar\omega_3 ) T_{03}, 
	\label{eq:momentum-mismatch2}
\end{align}
where $T_{\p\sigma} = L_1/v_\p - L_1/v_{\sigma}$ and $T_{0\sigma}  = L_2/v_{0} - L_2/v_{\sigma}$. Here, $v_\p = \partial k_\p / \partial \omega \vert_{\omega=\omega_\p}$ and $v_{\sigma} = \partial k_{\sigma} / \partial \omega \vert_{\omega=\bar\omega_{\sigma}}$ represent the group velocities of the input laser and the generated beam at the frequency $\bar\omega_\sigma$, respectively. Without loss of generality, we assume that $T_{\p0} \geq T_{\p1}$ and $T_{02} \ge T_{03}$. We merge all other constants into a factor, $\eta$, in Eq.~\eqref{eq:three-photon-amplitude}, which corresponds to the conversion efficiency of the cascaded PDC process.

In this study, we focus on monochromatic pumping with frequency $\omega_\p$ for the cascaded PDC process. In this situation, the energy conservation in the two processes is satisfied as $\omega_\p = \omega_1 + \omega_2 + \omega_3$. The three-photon amplitude in Eq.~\eqref{eq:three-photon-amplitude} can be rewritten as \cite{Shalm:2013ia}
\begin{align}
	f(\omega_1,\omega_2, \omega_3) 
	&= 
	\eta \delta(\omega_1+\omega_2+\omega_3-\omega_\p)
	r(\omega_1,\omega_3),
	\label{eq:monochromatic-three-photon}
\end{align}
where $r(\omega_1,\omega_3)=\phi(\omega_1, \omega_\p-\omega_1-\omega_3, \omega_3)$ is given by
\begin{align}
    r(\omega_1,\omega_3) 
    &= 
    \sinc\frac{(\omega_1 - \bar\omega_1)T_\en^{(01)}}{2}
\notag \\
    &\quad \times  
    \sinc\frac{(\omega_1 - \bar\omega_1)T_{02} + (\omega_3 - \bar\omega_3)T_\en^{(23)}}{2}.
	\label{eq:monochromatic-PMF}
\end{align}
The difference, $T_\en^{(01)} =  T_{\p0}-T_{\p1}$, is the entanglement time between photons~0 and 1, \cite{Saleh:1998vl} which represents the maximum relative delay between photons~0 and 1. Similarly, in the secondary PDC, the entanglement time between photons~2 and 3 is defined by $T_\en^{(23)} = T_{02}-T_{03}$.

\subsection{Frequency-dispersed two-photon coincidence counting measurement}

\begin{figure}
	\includegraphics{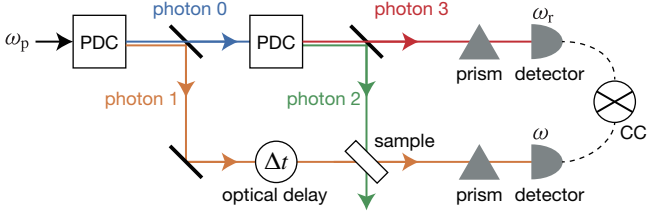}
	\caption{
	Schematic of frequency-dispersed two-photon coincidence counting measurement using entangled three photons generated via cascaded PDC pumped with a monochromatic laser of frequency $\omega_\p$. 
	Photons~1 and 2 are directed onto a sample with the external time of the photon~1 beam delay, $\dt$. Photon~3 does not interact with the sample and is detected in coincidence with photon~1, the latter of which is transmitted through the sample by the coincidence counter (CC).
	}
	\label{fig:1}
\end{figure}

We considered the frequency-dispersed two-photon coincidence counting measurement using the entangled three photons. The delay intervals between two of the three photons are innately determined when generated in the cascaded PDC. For example, the upper bound of the interval between photons~1 and 2 is given by $(T_\en^{(01)} + \lvert T_{02} \rvert )/2$. In a similar fashion to the two-photon interference experiments, \cite{Hong:1987gm,Franson:1989go} the delay intervals among the three photons can be controlled by adjusting the path differences between the beams. \cite{Agne:2017ob,Menssen2017:di} Therefore, the interval between photons~1 and 2 is experimentally controllable when the external time delay is sufficiently long compared to $T_\en^{(01)}$ and $T_{02}$. This external time delay is herein denoted as $\dt$. 
As presented in Fig.~\ref{fig:1}, photon~2 is employed as the pump field, whereas photon~1 is used for the probe field with the time delay $\dt \ge 0$. Photon~3 does not interact with the sample; it serves as a reference for the coincidence measurement. 
We assume that the efficiency of the photon detectors is perfect. In this situation, the detection of photon~3 makes it possible to verify the generation of entangled three photons.
Consequently, the coincidence measurements of photons~1 and 3 enable us to distinguish the genuine spectroscopic signal induced by two of the entangled three photons from undesired accidental photon counts in the photon~1 detector.
This is a potential benefit of utilizing two-photon coincidence detection to conduct measurements.

We consider a system comprising molecules and light fields. The positive-frequency component of the electric field operator, which interacts with the molecules, is written as \cite{Hong:1987gm,Franson:1989go,Ishizaki:2020jl} 
\begin{align}
	\E(t) 
	= 
	\E_1(t) 
	+ 
	\E_2(t+\dt), 
	\label{eq:field-operator}
\end{align}
where $\E_\sigma(t) = (2\pi)^{-1}\int d\omega\, \hat{a}_\sigma(\omega)  e^{-i \omega t}$. Here, the slowly varying envelope approximation has been adapted with the bandwidth of the fields assumed to be negligible in comparison to the central frequency.\cite{loudon2000quantum} Under the rotating-wave approximation, the molecule--field interaction can be written as $\hat{H}_\text{mol--field}(t) = -\hat{\mu}^\dagger \E(t) - \hat{\mu} \E^\dagger(t)$, where the dipole operator $\hat{\mu} $ is defined by $\hat{\mu} = \sum_\alpha \mu_{\alpha 0} \lvert 0  \rangle \langle e_\alpha \rvert + \sum_{\alpha \bar\gamma} \mu_{\bar\gamma \alpha} \lvert e_\alpha \rangle \langle f_{\bar\gamma} \rvert$.
In the above, $\lvert 0 \rangle $ represents the electronic ground state in the molecules. The summations are performed on indices that run over electronic excited states in the single-excitation manifold $\{ \lvert e_\alpha \rangle \}$ and double-excitation manifold $\{ \lvert f_{\bar\gamma} \rangle \}$. The probe fields transmitted through the sample, $\E_1$, and the reference field, $\E_3$, are both frequency-dispersed. Then, changes in the two-photon counting rate, $\tr [ \hat{a}_3^\dagger(\omega_{\rm r})  \hat{a}_1^\dagger(\omega) \hat{a}_1(\omega) \hat{a}_3(\omega_{\rm r}) \hat\rho(\infty) ]$, are measured. Thus, the frequency-dispersed two-photon counting signal is written as \cite{Dorfman:2016da,Schlawin:2017ea,Ye:2020dz}
\begin{align}
	S(\omega, \omega_{\rm r} ; \dt)
	&=
	{\rm Im} \int^\infty_{-\infty} dt \, e^{i\omega t}
\notag \\
	&\quad \times
	\tr[ \hat{a}_3^\dagger(\omega_{\rm r}) \hat{a}_3(\omega_{\rm r}) \E_1^\dagger(\omega) \hat\mu \hat\rho(t) ].
	\label{eq:transmission}
\end{align}
The initial conditions are: $\hat\rho(-\infty) = \lvert 0 \rangle\langle 0 \rvert \otimes \lvert \psi_\cpdc \rangle\langle \psi_\cpdc \rvert$.
The lowest-order contribution of Eq.~\eqref{eq:transmission} only comprises the absorption of photon~1.
However, the absorption signal is independent of the frequency of the input laser, $\omega_\p$, and the delay time, $\dt$, as shown in Appendix~C. 
Moreover, when the entanglement time $T_\en^{(23)}$ is much shorter than the characteristic timescales of the dynamics under investigation, the absorption signal is independent of the frequency of the input laser, $\omega_\p$, reference frequency, $\omega_\text{r}$, and the delay time, $\dt$.
Thus, in the two-photon coincidence measurement, which improves the signal-to-noise ratio, this process can be separated from the pump-probe-type two-photon process.
Consequently, the perturbative expansion of $\hat\rho(t)$ with respect to the molecule--field interaction, $\hat{H}_\text{mol--field}$, yields the third-order term as the leading order contribution.
The resultant signal is expressed as the sum of eight contributions, which are classified into stimulated emission (SE), ground-state bleaching (GSB), excited-state absorption (ESA), and double-quantum coherence (DQC).

\begin{figure}
	\includegraphics{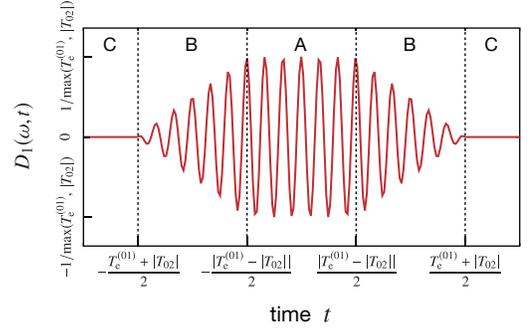}
	\caption{Plot of $D_1(\omega,t)$ from Eq.~\eqref{eq:Dn-setup1} as a function of $t$. 
	$D_1(\omega,t)$ is expressed as: 
	$D_1(\omega,t) =\exp[{i(\omega - \bar\omega_3)(T_\en^{(23)}/\lvert T_{02} \rvert)t}]/{\rm max}(T_\en^{(01)},\lvert T_{02} \rvert)$ in region A. 
	Furthermore, $D_1(\omega,t) = \exp[i(\omega - \bar\omega_3)(T_\en^{(23)}/\lvert T_{02} \rvert)t] (T_\en^{(01)}+\lvert T_{02} \rvert -2\lvert t \rvert )/(2 T_\en^{(01)} \lvert T_{02} \rvert)$ in region B, and $D_1(\omega,t)=0$ in region C.}
	\label{fig:2}
\end{figure}

To obtain a concrete but simple expression of the signal, here the memory effect straddling different time intervals in the response function is ignored.\cite{Ishizaki:2012kf} 
Hereafter, the reduced Planck constant, $\hbar$, is omitted.
Consequently, Eq.~\eqref{eq:transmission} can be expressed as (see Appendix~B)
\begin{align}
    S(\omega,\omega_{\rm r};\Delta t)
    &=
    S_\SE(\omega, \omega_{\rm r};\dt) 
    +
    S_\GSB(\omega, \omega_{\rm r};\dt) 
\notag \\
    &\quad +
    S_\ESA(\omega, \omega_{\rm r};\dt)
\notag \\
    &\quad +  
    S_\DQC(\omega, \omega_{\rm r};\dt)
	\label{eq:transmission-sum}
\end{align}
in terms of the SE, GSB, ESA, and DQC contributions,
\begin{align}
	S_\SE(\omega, \omega_{\rm r};\dt) 
	&=
	-
    {\rm Re}
	\sum_{\alpha\beta\gamma\delta} 
	\mu_{\gamma 0}
	\mu_{\delta 0}
	\mu_{\beta 0}
	\mu_{\alpha 0}
\notag \\ 
	&\quad \times
	\sum_{y={\rm r},{\rm nr}}
    I_{\gamma 0;\gamma\delta \gets \alpha\beta;\alpha 0}^{(y)}
	(\omega, \omega_{\rm r};\dt)
\notag \\
	&\quad
	+
	\sum_{y=\rm{r},\rm{nr}}
	\Delta S_\SE^{(y)}(\omega, \omega_{\rm r}),
	\label{eq:SE-general}
\end{align}
\begin{align}
	S_\GSB(\omega, \omega_{\rm r};\dt) 
	&=
	-
    {\rm Re}
	\sum_{\alpha\beta} 
	\mu_{\beta 0}^2
	\mu_{\alpha 0}^2
\notag \\ 
	&\quad \times
	\sum_{y=\rm{r},\rm{nr}}
	I_{\beta 0;00 \gets 00;\alpha 0}^{(y)}
	(\omega, \omega_{\rm r};\dt)
\notag \\
	&\quad
	+
	\sum_{y={\rm r},{\rm nr}}
	\Delta S_\GSB^{(y)}(\omega, \omega_{\rm r}),
	\label{eq:GSB-general}
\end{align}
\begin{align}
	S_\ESA(\omega, \omega_{\rm r};\dt) 
	&=
    +{\rm Re}
	\sum_{\alpha\beta \gamma\delta \bar\epsilon} 
	\mu_{\bar\epsilon \delta}
	\mu_{\bar\epsilon \gamma}
	\mu_{\beta 0}
	\mu_{\alpha 0}
\notag \\ 
	&\quad \times
	\sum_{y=\rm{r},\rm{nr}}
	I_{\bar\epsilon \delta;\gamma\delta \gets \alpha\beta; \alpha 0}^{(y)}
	(\omega, \omega_{\rm r};\dt),
	\label{eq:ESA-general}
\end{align}
\begin{align}
	S_\DQC(\omega, \omega_{\rm r} ; \dt)
	&=
	\mathrm{Re}
	\sum_{\alpha\beta\bar\gamma} 
	 \mu_{\bar\gamma  \beta} \mu_{\beta 0} \mu_{\bar\gamma \alpha} \mu_{\alpha 0}
\notag \\ 
	&\quad \times
	\left\{
	  G_{\bar\gamma\beta}[\omega] G_{\bar\gamma 0}[\omega_\p - \omega_{\rm r}] 
	  F'_{\alpha 0}(\omega, \omega_{\rm r};\dt)
	  \right.
\notag \\
	 & \left. \quad -
	  G_{\beta 0}[\omega] G_{\bar\gamma 0}[\omega_\p - \omega_{\rm r}] 
	  F'_{\alpha 0}(\omega, \omega_{\rm r};\dt)
	  \right\},
	\label{eq:DQC-general}
\end{align}
where $y$ indicates ``rephasing'' (r) or ``non-rephasing'' (nr), and $G_{\alpha\beta}(t)$ describes the time evolution of the $\lvert e_\alpha \rangle\langle e_\beta \rvert$ coherence. The Fourier-Laplace transform of $G_{\alpha\beta}(t)$ is introduced as $G_{\alpha\beta}[\omega] = \int^\infty_0 dt\,e^{i\omega t} G_{\alpha\beta}(t)$.
In Eqs.~\eqref{eq:SE-general} -- \eqref{eq:ESA-general}, the function $I_{\epsilon \zeta;\gamma\delta \gets \alpha\beta;\alpha 0}^{(y)}(\omega, \omega_{\rm r};\dt)$ is defined by 
\begin{align}
    I_{\epsilon \zeta;\gamma\delta \gets \alpha\beta; \alpha 0}^\r(\omega, \omega_{\rm r};\dt)
	&=	
	G_{\epsilon \zeta}[\omega]
    F_{\gamma\delta \gets \alpha\beta}(\omega, \omega_{\rm r};\dt, 0)
\notag \\ 
	&\quad \times   
    G_{\alpha 0}^\ast[\omega_\p -\omega_{\rm r} - \omega ],
    \label{eq:Ir-expression}
\end{align}
\begin{multline}
    I_{\epsilon \zeta;\gamma\delta \gets \alpha\beta;\alpha 0}^\nr(\omega, \omega_{\rm r};\dt)
	=		
	G_{\epsilon \zeta}[\omega]
	\int_0^\infty ds_1
	e^{i(\omega_\p -\omega_{\rm r} - \omega)  s_1}
\\
	\times
	F_{\gamma\delta \gets \alpha\beta}(\omega, \omega_{\rm r};\dt,s_1)
	G_{\alpha 0} (s_1)
	\label{eq:Inr-expression}
\end{multline}
in terms of
\begin{align}
	&F_{\gamma\delta \gets \alpha\beta}(\omega, \omega_{\rm r};\dt,s_1) 
\notag \\
	&\quad=
	r (\omega,\omega_{\rm r})
	\int_0^\infty ds_2
	G_{\gamma\delta \gets \alpha\beta}(s_2)
\notag \\ 
	&\quad\quad\times   
	e^{ -i (\omega - \bar\omega_1) \dt}
	[
        \D_1(\omega_{\rm r},s_2 + s_1 - \dt)
        e^{ i (\omega - \bar\omega_1) (s_2+s_1)}
\notag \\       
        &\quad\quad+
        \D_1(\omega_{\rm r},s_2 + s_1 + \dt)
        e^{ i (\omega +\omega_{\rm r} - \bar\omega_2 - \bar\omega_3) (s_2+s_1)}
        ],
	\label{eq:F-expression}
\end{align}
where $G_{\gamma\delta \gets \alpha\beta}(t)$ is the matrix element of the time-evolution operator defined by $\rho_{\gamma\delta}(t) = \sum_{\alpha\beta} G_{\gamma\delta \gets \alpha\beta}(t-s) \rho_{\alpha\beta}(s)$.
In Eq.~\eqref{eq:DQC-general}, $F'_{\alpha \beta}(\omega, \omega_{\rm r};\dt)$ is defined by
\begin{align}
	F'_{\alpha \beta}&(\omega, \omega_{\rm r};\dt) 
\notag \\
	&\quad=
	r (\omega,\omega_{\rm r})
	\int_0^\infty ds_1
	G_{\alpha \beta}(s_1)
\notag \\ 
	&\quad\quad\times   
	e^{ -i (\omega - \bar\omega_1) \dt}
	[
        \D_1(\omega_{\rm r},s_1 + \dt)
        e^{ i \bar\omega_1 s_1}
\notag \\ 
	&\quad\quad    
        +
        \D_1(\omega_{\rm r},s_1 - \dt)
        e^{ i  (\bar\omega_2+\bar\omega_3- \omega_{\rm r}) s_1}
        ],
	\label{eq:F-DQC-expression}
\end{align}
The function $\D_n(\omega, t)$ $(n=1,2,\dots)$ is introduced as
\begin{align}
    \D_n(\omega, t)
    =
    \int^\infty_{-\infty}
	\frac{d\xi}{2\pi}
	e^{-i\xi t} 
	r(\xi + \bar\omega_1,\omega)^n .
	\label{eq:Dn-setup1}
\end{align}
Note that $D_1(\omega_{\rm r}, t)$ is non-zero when $\lvert t \rvert \le (T_\en^{(01)} + \lvert T_{02} \rvert  )/2$, as illustrated in Fig.~\ref{fig:2}. The $\dt$-independent term in Eqs.~\eqref{eq:SE-general} and \eqref{eq:GSB-general}, $\Delta S_{x}^{(y)}(\omega, \omega_{\rm r})$, originates from the field commutator. Details of the $\dt$-independent terms are given in Appendix~D.

To understand the influence of entanglement times on the spectrum in Eq.~\eqref{eq:transmission-sum}, here we investigate the limiting cases.
When $T_\en^{(01)}$, $T_\en^{(23)}$, and $T_{02}$ are much shorter than the characteristic timescales of the dynamics under investigation,\cite{Ishizaki:2020jl,Fujihashi:2020ep} we approximately obtain $r(\omega_1,\omega_3) = 1$ and $\D_n(\omega, t) = \delta(t)$. Consequently, Eqs.~\eqref{eq:F-expression} and \eqref{eq:F-DQC-expression} can be simplified as
\begin{align}
	F_{\gamma\delta \gets \alpha\beta}(\omega, \omega_{\rm r};\dt,s) 
	&= 
	G_{\gamma\delta \gets \alpha\beta}(\dt-s),
	\label{eq:F-expression-limit}
\\[12pt]
	F'_{\alpha \beta}(\omega, \omega_{\rm r};\dt) 
	&=
	G_{\alpha \beta}(\dt)
	e^{ i (\omega_p - \omega - \omega_{\rm r}) \dt},
\end{align}
and thus, $I_{\epsilon \zeta;\gamma\delta \gets \alpha\beta;\alpha 0}^{(y)}(\omega, \omega_{\rm r};\dt)$ is written as
\begin{multline}
    I_{\epsilon \zeta;\gamma\delta \gets \alpha\beta; \alpha 0}^{(y)}(\omega, \omega_{\rm r};\dt)
\\
	=
	G_{\epsilon \zeta}[\omega]
	G_{\gamma\delta \gets \alpha\beta}(\dt)
	G_{\alpha 0}^{(y)}[\omega_\p -\omega_{\rm r} - \omega ],
	\label{eq:r-limit}
\end{multline}
where $G_{\alpha 0}^\r[\omega] = G_{\alpha 0}^\ast[\omega]$ and $G_{\alpha 0}^\nr[\omega] = G_{\alpha 0}[\omega]$ have been introduced. In deriving Eq.~\eqref{eq:r-limit}, we assume that $G_{\gamma\delta \gets \alpha\beta}(\dt-s_1) G_{\alpha 0}(s_1) \simeq G_{\gamma\delta \gets \alpha\beta}(\dt) G_{\alpha 0}(s_1)$ in the non-rephasing case.\cite{Cervetto:2004gm} This approximation is justified when the response function varies slowly as a function of the waiting time, $\dt$. As was demonstrated in Ref.~\onlinecite{Ishizaki:2020jl}, the signal $S(\omega,\omega_{\rm r};\dt)$ corresponds to the spectral information along the anti-diagonal line, $\omega_1 + \omega_3 = \omega_\p - \omega_{\rm r}$, on the absorptive two-dimensional (2D) spectrum $\mathcal{S}_{\rm 2D}(\omega_3, t_2, \omega_1)$,
\begin{align}
	S(\omega,\omega_{\rm r};\dt) 
	\simeq
	-
	\mathcal{S}_{\rm 2D}(\omega, \dt, \omega_\p - \omega_{\rm r} - \omega),
	\label{eq:correspondence}
\end{align}
except for the $\dt$-independent terms in Eqs.~\eqref{eq:SE-dt0-limit} and \eqref{eq:GSB-dt0-limit}, respectively. Equation~\eqref{eq:correspondence} indicates that the two-photon counting signal $S(\omega,\omega_{\rm r};\dt)$, is homologous to the 2D spectrum, $\mathcal{S}_{\rm 2D}(\omega_3, \dt,\omega_1)$. This is true even when the frequency of the input laser, $\omega_\p$, is fixed. This correspondence is similar to, but different from, the results reported by Ref.~\onlinecite{Ishizaki:2020jl}, wherein the transmission signal was found to provide the same information as the 2D spectrum only when sweeping the frequency of the input laser, $\omega_\p$. In addition, we consider the opposite limit, $T_\en^{(01)} \to \infty$ and $T_\en^{(23)} \to \infty$. We obtain $r(\omega_1,\omega_3) = \delta(\omega_1-\bar\omega_1) \delta(\omega_3-\bar\omega_3)$. Equations~\eqref{eq:F-expression} and \eqref{eq:F-DQC-expression} can thus be written as 
\begin{multline}
    F_{\gamma\delta\gets\alpha\beta}(\omega,\omega_{\rm r},\dt, s)
\\  
    \propto
    \delta(\omega-\bar\omega_1)
    \delta(\omega_{\rm r}-\bar\omega_3)
    G_{\gamma\delta \gets \alpha\beta}[0],
\end{multline}
\begin{multline}
	F'_{\alpha \beta}(\omega, \omega_{\rm r};\dt)
\\	
	\propto
    \delta(\omega-\bar\omega_1)
    \delta(\omega_{\rm r}-\bar\omega_3)
    \left(
    G_{\alpha \beta}[\bar\omega_1]
    +
    G_{\alpha \beta}[\bar\omega_2]
    \right),
\end{multline}
where $G_{\gamma\delta \gets \alpha\beta}[0] = \int_0^\infty dt\,G_{\gamma\delta \gets \alpha\beta}(t)$ is defined. In this limit, the temporal resolution is eroded, and the spectrum in Eq.~\eqref{eq:transmission-sum} does not provide any information on the excited-state dynamics.

\section{Numerical results and discussion}
\begin{figure}
	\includegraphics{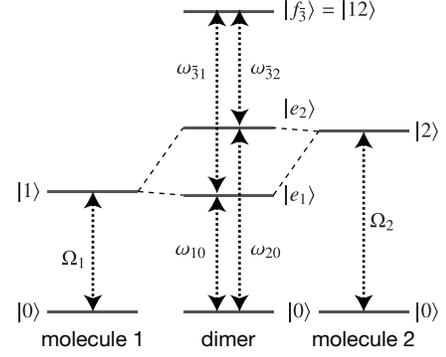}
	\caption{
	Illustration of quantum superposition between the electronic transitions of molecules 1 and 2 in a coupled dimer.
	}
	\label{fig:3}
\end{figure}

\setcounter{equation}{0}
To numerically demonstrate Eq.~\eqref{eq:transmission-sum} using Eqs.~\eqref{eq:SE-general} -- \eqref{eq:Dn-setup1}, we consider the electronic excitations in a coupled dimer, as depicted in Fig.~\ref{fig:3}. The electronic excitation Hamiltonian is expressed as $\hat{H}_{\rm ex} = \sum_m \hbar\Omega_m \hat{B}_m^\dagger \hat{B}_m + \sum_{m\ne n} \hbar J_{mn}\hat{B}_m^\dagger \hat{B}_n$, where $\hbar\Omega_m$ is the Franck-Condon transition energy of the $m$-th molecule and $\hbar J_{mn}$ is the electronic coupling between the $m$-th and $n$-th molecules. \cite{Ishizaki:2012kf} 
In the Hamiltonian, the excitation creation operator $\hat{B}_m^\dagger$ is introduced for the excitation vacuum $\lvert 0 \rangle$, such that $\lvert m \rangle = \hat{B}_m^\dagger \lvert 0 \rangle$ and $\lvert mn \rangle = \hat{B}_{m}^\dagger\hat{B}_n^\dagger \lvert 0 \rangle$. 
In the eigenstate representation, the excitation Hamiltonian can be written as $\hat{H}_{\rm ex} = \epsilon_0 \lvert 0 \rangle \langle 0 \rvert + \sum_\alpha \epsilon_\alpha \lvert e_\alpha \rangle \langle e_\alpha \rvert + \sum_{\bar{\gamma}} \epsilon_{\bar{\gamma}} \lvert f_{\bar{\gamma}} \rangle \langle f_{\bar{\gamma}} \rvert$, where $\lvert e_\alpha \rangle = \sum_m V_{m \alpha} \lvert m \rangle$ and $\lvert f_{\bar{\gamma}} \rangle = \sum_{mn} W_{m n, \bar{\gamma}} \lvert mn \rangle$. Accordingly, the exciton transition dipole moments are expressed as $\mu_{\alpha 0}= \sum_m V_{\alpha m}^{-1} \mu_{m 0} $ and $\mu_{ \bar{\gamma} \alpha}= \sum_{mn} W_{\bar{\gamma} (mn)}^{-1} V_{\alpha m}^{-1} \mu_{n0}$. 
We assume that the environmentally-induced fluctuations in the electronic energies are described as a Gaussian process. By applying the second-order cumulant expansion for the fluctuations, the third-order response function is expressed in terms of the line-broadening function, $g_m(t)=\int_0^t ds_1 \int_0^{s_1} ds_2 C_m (s_2)$, where $C_m (t)$ is expressed as $C_m (t)=\int^\infty_0 d\omega J_m(\omega)[\coth(\hbar \omega/2k_{\rm B}T)\cos \omega t - i \sin \omega t ]$ in terms of the spectral density, $J_m(\omega)$.
In this study, the spectral density is modeled as $J_m(\omega)= 8 E_{\rm env} \gamma_{\rm env}^3 \omega / (4 \omega^2 + \gamma_{\rm env}^2 )^2$, where $E_{\rm env}$ and $\gamma_{\rm env}^{-1}$ represent the energy and timescale of the environmental reorganization, respectively. \cite{Ishizaki:2020jy} 
To describe the time-evolution of the electronic excitations in the waiting time, the electronic coherence in the single excitation manifold is ignored, and hence, $G_{\beta\beta \gets \alpha\alpha}(t)$ in Eq.~\eqref{eq:F-expression} is computed with the master equation, 
\begin{align}
    \frac{d}{dt}G_{\beta\beta \gets \alpha\alpha}(t)
    &=
    \sum_{\xi (\ne \beta)} 
    k_{\beta \gets \xi} G_{\xi\xi \gets \alpha\alpha}(t) 
\notag \\   
    &\quad
    -
    \sum_{\xi (\ne \beta)} 
    k_{\xi \gets \beta} G_{\beta\beta \gets \alpha\alpha}(t), 
	\label{eq:MRT}
\end{align}
where the rate constant $k_{\beta \gets \alpha}$ is obtained with the modified Redfield theory.\cite{Zhang:1998eo,Yang:2002ik}
With the initial condition of $G_{\beta\beta \gets \alpha\alpha}(0) = \delta_{\beta\alpha}$, the equation leads to
\begin{align}
    G_{\beta\beta \gets \alpha\alpha}(t)
    =
    \sum_\xi 
    g_{\beta\alpha}^{(\xi)}
    e^{-\lambda_\xi t},
	\label{eq:MRT2}
\end{align}
with $g_{\beta\alpha}^{(\xi)} =  U_{\beta \xi} (U^{-1})_{\xi \alpha}$,
where $\lambda_\xi$ is the $\xi$-th eigenvalue of the matrix whose element is $K_{\xi\xi'} = \delta_{\xi\xi'}\sum_{\gamma (\neq \xi)} k_{\gamma \gets \xi} + (1-\delta_{\xi\xi'})k_{\xi \gets \xi'}$, and $U_{\alpha \xi}$ is an element of the modal matrix as such $\lambda_{\xi} = (U^{-1} K U)_{\xi\xi}$.

For numerical calculations, we set the gap between the Franck-Condon transition energies of pigments~1 and 2 to $\Omega_2 -\Omega_1= 200\cm$. Furthermore, we set their electronic coupling to $J_{12}=50\cm$.
For simplicity, we set the transition dipole strengths as $\mu_{10}=\mu_{20}=1$.
We set the reorganization energy, relaxation time, and temperature as $E_{\rm env}=35\cm$, $\gamma_{\rm env}^{-1}=50\fs$, and $T=77\,{\rm K}$, respectively.
Under this condition, the energy gap between the eigenstates, $\omega_{20} - \omega_{10} = 224\cm$, is much higher than the thermal energy. Therefore, the influence of the uphill excitation transfer, $e_1 \to e_2$, on the signal can be considered to be small.

\subsection{Limit of short entanglement time}

\begin{figure}
	\includegraphics{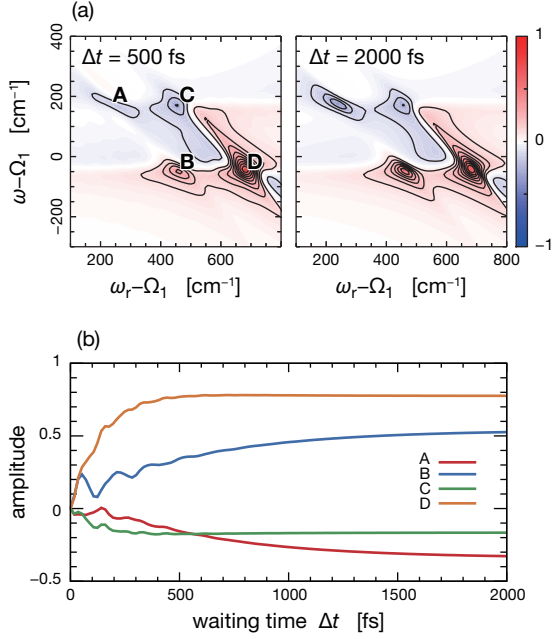}
	\caption{
	(a) Difference spectra $\Delta S (\omega,\omega_{\rm r}; \dt) = S(\omega,\omega_{\rm r};\dt) - S(\omega,\omega_{\rm r};0)$ for $T_\en^{(01)} =T_\en^{(23)} = T_{02} =  0$.
	The waiting times are $\dt=500\fs$ and $2000\fs$.
	The frequency of the input laser is fixed as $\omega_\p=3\Omega_2$.	
	The parameters in the dimer systems are $\Omega_2-\Omega_1 = 200\cm$, $J_{12}=50\cm$, $E_{\rm env}=35\cm$, $\gamma_{\rm env}^{-1}=50\fs$, and $T=77\,{\rm K}$. 
	The normalization of the contour plot is such that the maximum value of the spectrum is unity, and equally-spaced contour levels ($\pm 0.1$, $\pm 0.2$, \dots) are drawn.
	Panel~(b) shows the amplitudes of peak~A ($\omega_{\rm r}-\Omega_1=258\cm$, $\omega-\Omega_1=171\cm$), peak~B ($\omega_{\rm r}-\Omega_1=479\cm$, $\omega-\Omega_1=-50\cm$), peak~C ($\omega_{\rm r}-\Omega_1=479\cm$, $\omega-\Omega_1=171\cm$), and peak~D ($\omega_{\rm r}-\Omega_1=700\cm$, $\omega-\Omega_1=-50\cm$) as a function of the waiting time, $\dt$.
	}
	\label{fig:4}
\end{figure}

To demonstrate how the spectrum provides time-resolved information on the state-to-state dynamics, we first investigate the limit of the short entanglement time, $T_\en^{(01)} = T_\en^{(23)} = T_{02} = 0$, although it may be unrealistic for the cascaded PDC process.
As shown in Eqs.~\eqref{eq:SE-general} and \eqref{eq:GSB-general}, the spectrum contains $\dt$-independent contributions. Therefore, we consider the difference spectrum,
\begin{align}
    \Delta S(\omega,\omega_{\rm r};\dt)
    =
    S(\omega,\omega_{\rm r};\dt) - S(\omega,\omega_{\rm r};0).
	\label{eq:difference-spectrum}
\end{align}
Figure~\ref{fig:4}(a) presents the difference spectra of the model dimer for two different waiting times, $\dt$, when the frequency of the input laser is $\omega_\p=3\Omega_2$.
The waiting times are $\dt=500\fs$ and $2000\fs$.
Figure~\ref{fig:4}(a) shows strong signatures of the ESA signal at the location A, and strong signatures of the SE signal at the location labeled B.
As was clarified in Eq.~\eqref{eq:r-limit}, the possible pairs of optical transitions probed at frequency $\omega=\omega_{\epsilon\zeta}$ are restricted by the resonance condition, $\omega_{\alpha 0}+\omega_{\epsilon \zeta}+\omega_{\rm r} \simeq \omega_\p$, which is imposed by the non-classical correlations among the entangled three photons.
Hence, the negative peak at A corresponds to the pair of optical transitions ($0 \to e_2$, $e_1 \to f_{\bar{3}}$), while the positive peak at B corresponds to the pair of optical transitions ($0 \to e_2$, $e_1 \to 0$).
The increases in these signal amplitudes during the waiting period $\dt$ reflect the excitation relaxation $e_2 \to e_1$, as shown in Fig.~\ref{fig:4}(b).
Therefore, the two-photon counting signal temporally resolves the excitation relaxation $e_2 \to e_1$ through the changes in the amplitudes of peaks~A or B during the waiting period, $\dt$.

In Fig.~\ref{fig:4}(a), strong ESA and SE signals can also be observed at locations C and D, respectively.
These ESA and SE signals correspond to the pairs of optical transitions ($0 \to e_1$, $e_1 \to f_{\bar{3}}$) and ($0 \to e_1$, $e_1 \to 0$), respectively.
As shown in Fig.~\ref{fig:4}(b), the difference spectrum exhibited changes in the amplitudes of peaks~C and ~D occurring within $500\fs$; these peaks are much faster than the excitation relaxation, $e_2 \to e_1$.
Moreover, Fig.~\ref{fig:4}(b) exhibits the oscillatory transients of peaks~A and B, which persisted up to $\dt<500\fs$. However, the electronic coherence in the single-excitation manifold is not considered in this instance.
To understand these transient behaviors, we consider the non-rephasing contribution of the ESA signal in Eq.~\eqref{eq:Inr-expression}. 
For demonstration purposes, we assume that the time evolution in the $t_1$ period is denoted by $G_{\alpha 0}(t_1)=e^{-(i\omega_{\alpha 0}+\Gamma_{\alpha 0})t_1}$.
With the use of Eqs.~\eqref{eq:F-expression-limit} and \eqref{eq:MRT2}, the expression of $I_{\bar\epsilon \beta;\beta\beta \gets \alpha\alpha;\alpha 0}^\nr(\omega, \omega_{\rm r};\dt)$ in Eq.~\eqref{eq:Inr-expression} can be expressed as
\begin{multline}
    I_{\bar\epsilon \beta;\beta\beta \gets \alpha\alpha;\alpha 0}^\nr
    (\omega, \omega_{\rm r};\dt)
\\
	=
	-i
	G_{\bar\epsilon \beta}[\omega]
	\sum_{\xi=1,2}
	g_{\beta\alpha}^{(\xi)}
	\frac{e^{i {\Delta\omega_{\alpha 0}} \dt -\Gamma_{\alpha 0}\dt}  -e^{-\lambda_\xi \dt} }{{\Delta\omega_{\alpha 0}} +i  (\Gamma_{\alpha 0}-\lambda_\xi)},
	\label{eq:Inr-expression1}
\end{multline}
where $\lambda_1= 0$, $\lambda_2= k_{1 \gets 2} + k_{2 \gets 1}$, and $\Delta\omega_{\alpha 0}=\omega_\p -\omega_{\rm r} - \omega -\omega_{\alpha 0}$.
Equation~\eqref{eq:Inr-expression1} demonstrates that the amplitude of peak~A oscillates at the frequency $\Delta\omega_{20}$. This is the detuning of the $0 \to e_2$ transition from the frequency of photon~2.
Similarly, the transient dynamics in peak~C reflect the decay of the $\lvert e_1 \rangle \langle 0 \rvert$ coherence.
Therefore, the transient dynamics in peaks~A and C are not directly related to the dynamics in the single-excitation manifold during the $t_2$ period.
The SE contributions to peaks~B and D in the short-time region can also be understood in the same manner.
If coherence $\lvert e_\alpha \rangle \langle e_\beta \rvert$ is considered, the time-evolution operator is modeled as $G_{\alpha\beta \gets \alpha\beta}(t_2) = e^{-(i \omega_{\alpha\beta} + \Gamma_{\alpha\beta})t_2}$. Thus, Eq.~\eqref{eq:Inr-expression} yields
\begin{multline}
    I_{\bar\epsilon \beta;\alpha\beta \gets \alpha\beta;\alpha 0}^\nr(\omega, \omega_{\rm r};\dt)
\\
	=
	-i	
	G_{\bar\epsilon \beta}[\omega]
	\frac{e^{i {\Delta\omega_{\beta 0}}\dt -\Gamma_{\alpha 0}\dt} - e^{-(i \omega_{\alpha\beta} + \Gamma_{\alpha\beta})\dt}}{{\Delta\omega_{\beta 0}} + i(\Gamma_{\alpha 0}-\Gamma_{\alpha\beta})}.
	\label{eq:Inr-expression2}
\end{multline}
Equation~\eqref{eq:Inr-expression2} includes the oscillating component at the detuning frequency $\Delta\omega_{\beta 0}$, as well as the oscillation originating from the $\lvert e_\alpha \rangle \langle  e_\beta \rvert$ coherence.
In complex molecular systems such as photosynthetic light-harvesting proteins, the lifetime of the electronic coherence is typically a few hundred femtoseconds. On this time scale, the contribution of the $\lvert e_\alpha \rangle \langle 0 \rvert$ coherence during the $t_1$ period to the signal in Eq.~\eqref{eq:Inr-expression2} cannot be ignored. 
In this respect, Eq.~\eqref{eq:Inr-expression2} indicates that it is difficult to extract relevant information on the electronic coherence from the oscillatory dynamics in the signal.

\subsection{Cases of finite entanglement times}

\begin{figure}
	\includegraphics{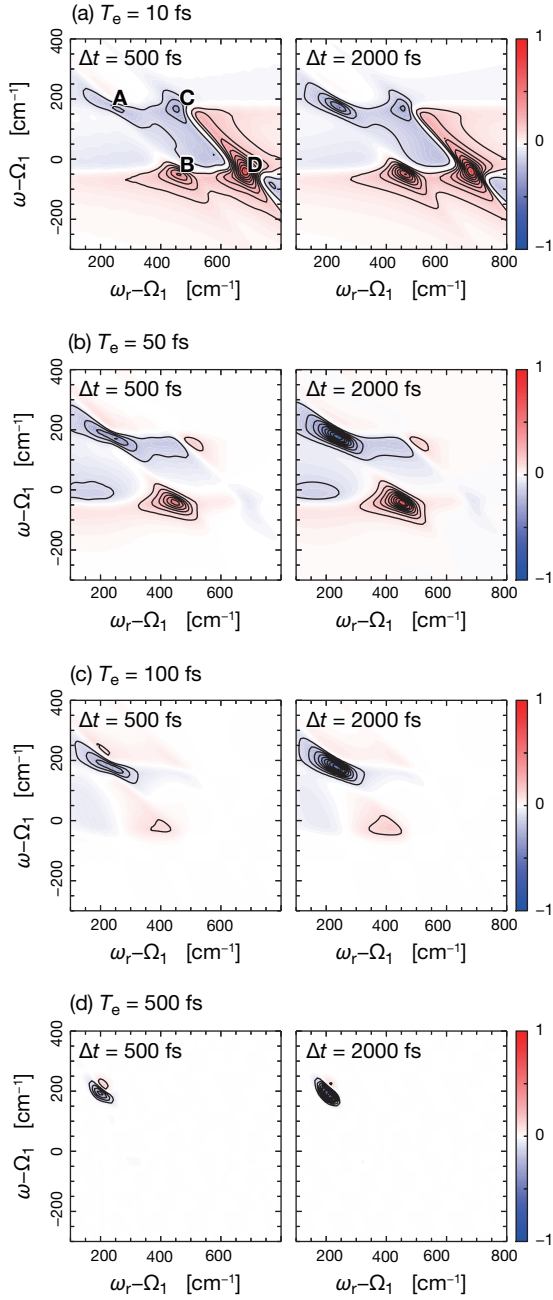}
	\caption{
	Difference spectra $\Delta S(\omega,\omega_{\rm r}; \dt) = S(\omega,\omega_{\rm r}; \dt) - S(\omega,\omega_{\rm r}; 0)$ for various values of the entanglement time, $T_\en \equiv T_\en^{(01)} =T_\en^{(23)} = 2T_{02}$: (a) $T_\en = 10\fs$, (b) $T_\en = 50\fs$, (c) $T_\en = 100\fs$, and (d) $T_\en = 500\fs$.
	The waiting times are $\dt=500\fs$ and $2000\fs$.
	The central frequencies of photons ~1, ~2, and ~3 are $\bar\omega_1=\bar\omega_2=\bar\omega_3=\omega_\p/3=\Omega_2$.
	The other parameters are the same as in Fig.~\ref{fig:4}. 
    }
	\label{fig:5}
\end{figure}

\begin{figure}
	\includegraphics{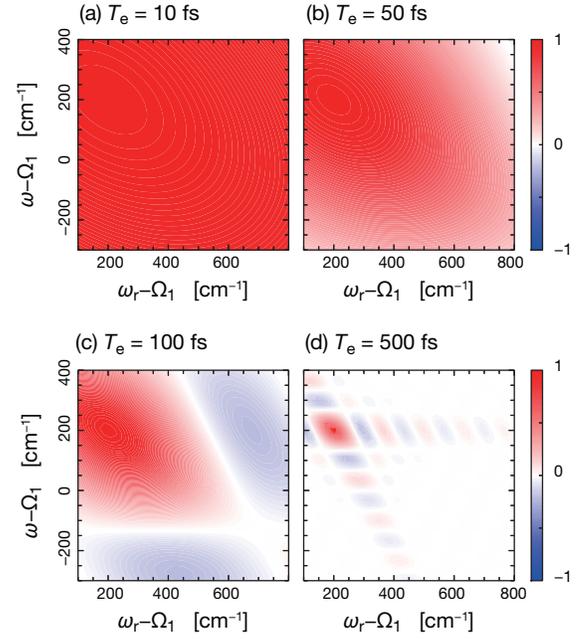}
	\caption{
	The phase-matching function $r(\omega,\omega_{\rm r})$ in Eq.~\eqref{eq:monochromatic-PMF} for the cases of (a) $T_\en = 10\fs$, (b) $T_\en = 50\fs$, (c) $T_\en = 100\fs$, and (d) $T_\en = 500\fs$.
	We set the central frequencies of the entangled three photons to $\bar\omega_1=\bar\omega_2=\bar\omega_3=\omega_\p/3=\Omega_2$.
	}
	\label{fig:6}
\end{figure}

\begin{figure}
	\includegraphics{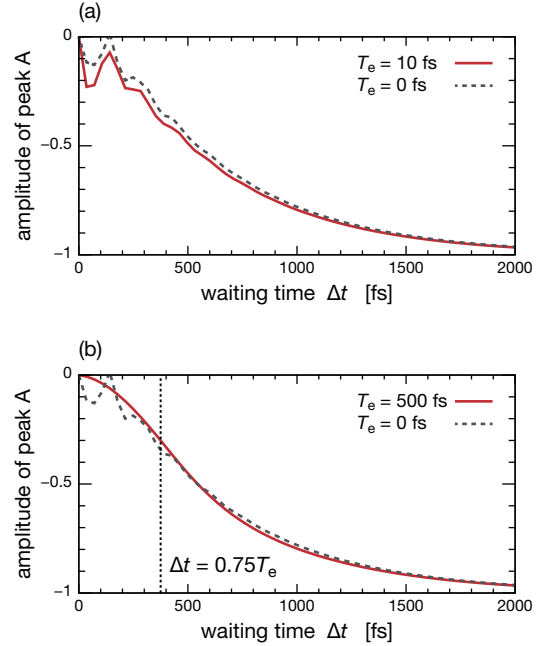}
	\caption{
	Time evolution of the amplitude of (a) peak~A ($\omega_{\rm r}-\Omega_1=258\cm$, $\omega-\Omega_1=171\cm$) in the case of $T_\en = 10\fs$ and (b) peak~A ($\omega_{\rm r} - \Omega_1=198\cm$, $\omega - \Omega_1=198\cm$) in the case of $T_\en = 500\fs$.
	In both panels, the gray dashed line shows the amplitude of peak~A ($\omega_{\rm r}-\Omega_1=258\cm$, $\omega-\Omega_1=171\cm$) in the limit of $T_\en = 0\fs$ as a reference.
	The normalization of the plots is such that the maximum value of peak~A is unity.
    }
	\label{fig:7}
\end{figure}

We investigate the effects of finite entanglement times on the spectrum. 
The central frequencies of the entangled three photons that have been generated can be varied by tuning the phase-matching conditions for the two PDC processes. \cite{Krapick:2016ch,Antonosyan:2011th} 
Therefore, we set the three central frequencies of the entangled three photons, that is, $\bar\omega_1 = \bar\omega_2 = \bar\omega_3 = \omega_\p/3 = \Omega_2$, which nearly resonate with the $0 \to e_2$ transition.
It is noted that $T_\en^{(01)}$, $T_\en^{(23)}$, and $T_{02}$ are not independent because the group velocities $v_\sigma = \partial k_{\sigma} / \partial \omega \vert_{\omega=\bar\omega_{\sigma}}$ depend on the central frequencies of the three photons, as presented in Eqs.~\eqref{eq:momentum-mismatch1} and \eqref{eq:momentum-mismatch2}.
For simplicity, we consider cases that satisfy $T_\en \equiv T_\en^{(01)} = T_\en^{(23)} = 2T_{02}$ in what follows. This condition corresponds to the equalities of $\partial k_{1} / \partial \omega \vert_{\omega=\omega_\p/3} = -  \partial k_{0} / \partial \omega \vert_{\omega=\omega_\p/3}$ and $ \partial k_{2} / \partial \omega \vert_{\omega=\omega_\p/3} = -  \partial k_{3} / \partial \omega \vert_{\omega=\omega_\p/3}$. In practical perspective, this requires us to explore the design of quasi-phase-matched crystals that can hold the equalities of the group velocities, which is beyond the scope of the present paper.

Figure~\ref{fig:5} presents the difference spectra of the molecular dimer in the cases of
(a) $T_\en = 10\fs$, (b) $T_\en =50\fs$, (c) $T_\en = 100\fs$, and (d) $T_\en = 500\fs$.
The other parameters are the same as those shown in Fig.~\ref{fig:4}.
The signal in Fig.~\ref{fig:5}(a) appears to be identical to the signal obtained under the three photon state in the limit of $T_\en = 0$, illustrated in Fig.~\ref{fig:4}(a).
However, the intensities of peaks~B, C, and D decrease with increasing entanglement time, $T_\en$, as presented in Figs.~\ref{fig:5}(b) -- (d), respectively.
In the case of $T_\en = 500\fs$, peaks~B, C, and D almost disappear.
To understand this dependence on the entanglement time, the rephasing contribution in Eq.~\eqref{eq:F-expression} is considered as an example.
Here, we note that $D_1(\omega_{\rm r}, t)$ is non-zero when $\lvert t \rvert \le 0.75\, T_\en $, as shown in Fig.~\ref{fig:2}.
In the case of $\dt > 0.75\, T_\en$, the expression of $F_{\beta\beta \gets \alpha\alpha}(\omega, \omega_{\rm r};\dt,0)$ is obtained as
\begin{multline}
	F_{\beta\beta \gets \alpha\alpha}(\omega, \omega_{\rm r};\dt,0)  
\\
	=
	r(\omega,\omega_{\rm r})
	\sum_{\xi=1,2}
	r(\omega +i\lambda_{\xi},\omega_{\rm r})
	g_{\beta\alpha}^{(\xi)}
	e^{-\lambda_\xi \dt}.
	\label{eq:F-expression1}
\end{multline}
The bandwidth of the phase-matching function in Eq.~\eqref{eq:monochromatic-PMF} is related to the inverse of the entanglement time, $T_\en$.
Equation~\eqref{eq:F-expression1} indicates that the finite entanglement time acts as a frequency filter through the spectral distribution of the phase-matching function, which limits the accessible spectral range.
Figure~\ref{fig:6} presents the spectral distribution of the phase-matching function in Eq.~\eqref{eq:monochromatic-PMF}.
Comparing Figs.~\ref{fig:5} and \ref{fig:6} reveals that all optical transitions that are outside the bandwidth of the phase-matching function are suppressed.
Therefore, the finite entanglement times can be used to selectively enhance specific Liouville pathways when the center frequencies of the entangled three photons are tuned to resonate with certain optical transitions. 
It is noteworthy that a similar property in terms of the finite entanglement time was discussed in the context of entangled two-photon spectroscopy.\cite{Munkhbaatar:2017se}

Further, we investigate the time-evolution of peak~A observed in the difference spectra (illustrated in Fig.~\ref{fig:5}).
In the case of $\dt > 0.75\, T_\en$, the contribution of the ESA signal at peak~A in Eq.~\eqref{eq:F-expression1} is written as
\begin{align}
	F_{11 \gets 22}
	( \bar\omega_1, \bar\omega_3; \dt,0)
	=
    g_{12}^{(1)}   
    + \Lambda g_{12}^{(2)} e^{-\lambda_2 \dt}
	\label{eq:F-expression2}
\end{align}
with $\Lambda = 8( \lambda_2 T_\en )^{-2} \sinh({ \lambda_2 T_\en}/{2}) \sinh({ \lambda_2 T_\en}/{4})$, where the approximations of $\bar\omega_1 \simeq \omega_{\bar{3}1}$ and $\bar\omega_3 \simeq \omega_\p - \omega_{\bar{3}1} -\omega_{20}$ are employed.
The $\dt$-dependence of Eq.~\eqref{eq:F-expression2} reflects the monotonous decay governed by the rate of the excitation transfer $\lambda_2= k_{1 \gets 2} + k_{2 \gets 1}$.
Therefore, the signal provides information on the dynamics of $e_2 \to e_1$ when $\dt > 0.75\, T_\en$, as shown in Fig.~\ref{fig:7}(b).
When the entanglement time, $T_\en$, is sufficiently short compared to the timescales of the excited-state dynamics, Eq.~\eqref{eq:F-expression2} becomes
$F_{11 \gets 22}(\bar\omega_1, \bar\omega_3; \dt,0)  \simeq
G_{11 \gets 22}(\dt)$, as presented in Fig.~\ref{fig:7}(a).
In contrast, in the case of $\dt < 0.75\, T_\en$, Eq.~\eqref{eq:F-expression} becomes 
\begin{multline}
	F_{11 \gets 22}
	(\bar\omega_1, \bar\omega_3; \dt,0)
	=
	- g_{12}^{(1)}
	- \frac{2 g_{12}^{(2)} }{\lambda_2^2 T_\en^2 }
	e^{-\lambda_2 T_\en}
\\
	 \times
	[ (1+\lambda_2 \dt) e^{-\lambda_2 \dt} 
	+ (1-\lambda_2 \dt) e^{\lambda_2 \dt}].
	\label{eq:F-expression3}
\end{multline}
Equation~\eqref{eq:F-expression3} demonstrates the complicated time-evolution, making it impossible to extract relevant information on the excited-state dynamics from the signal.
This can clearly be seen in Fig.~\ref{fig:7}(b), where it is not possible to temporally resolve the fast oscillatory transients within $0.75\, T_\en =375\fs$.
Figures~\ref{fig:5} -- \ref{fig:7} suggest that the manipulation of the phase-matching function enables filtering out a specific frequency region of the spectra while maintaining ultrafast temporal resolution, resulting in the achievement of the joint temporal and frequency resolution.

However, it is noted that the spectral and temporal resolution of the signal depends on the spectral shape of the entangled three photons.
For example, replacing the phase-matching function in Eq.~\eqref{eq:monochromatic-PMF} with a Gaussian distribution of the same bandwidth can improve the temporal resolution, as discussed in Appendix~E.

\section{Concluding remarks}
\setcounter{equation}{0}

The time-resolved spectroscopic measurement using the entangled photon pairs investigated in the preceding study \cite{Ishizaki:2020jl} faces the challenge in that it is difficult to separate the weak nonlinear optical signals from the linear absorption signal. In this work, we theoretically investigated the time-resolved spectroscopy utilizing entangled three photons generated via the cascaded PDC to overcome this difficulty. In this measurement, time-resolved spectroscopy with monochromatic pumping was integrated with the two-photon counting technique, which suppresses the undesired accidental photon counts in the detector and thus allows one to separate the weak nonlinear optical components from the remaining signals. It was also demonstrated that the frequency-dispersed two-photon counting signal provides the same spectral information as in a coherent 2D optical spectrum that requires the control of multiple laser pulses.
Furthermore, we investigated the influence of the finite entanglement times on the two-photon counting signal. The spectral distribution of the phase-matching function acts as a frequency filter to selectively resolve a specific region of the 2D spectrum, while the excited state dynamics under investigation are temporally resolved in a time domain that is longer than the entanglement time. This results in the achievement of the joint temporal and frequency resolution. It is thus anticipated that the time-resolved spectroscopy using the entangled three-photon system may be useful for investigating the dynamical processes in complex molecular systems, such as photosystem II reaction center, in which multiple electronic states are present within a narrow energy region.\cite{Fuller:2014iz,Romero:2014jm,Fujihashi:2015kz,Fujihashi:2018in}
However, it is still necessary to address several practical challenges in implementing the proposed spectroscopic scheme. The first issue is the low efficiency of three-photon generation via the cascaded PDC process, as pointed out in Sec.~IIA. Second, the performance of the coincidence measurement is very sensitive to the efficiency of the photon detector. \cite{Okamoto:2020lo} These issues could be overcome by devising a new entangled three-photon source, \cite{Corona:2011ex,Krapick:2016ch,Moebius:2016ef,Okoth:2019se,Dominguez:2020th} and by using the double-crystal interference technique, \cite{Lemos:2014qu,Kalashnikov:2016cl,Paterova:2018me} which does not require detection of photons transmitted through the sample. The extensions of the present work in these directions are to be explored in future studies.

\begin{acknowledgements}
YF would like to thank Ryosuke Shimizu for his valuable comments on this manuscript.
This work was supported by JST PRESTO Grant No.~JPMJPR19G8, JSPS KAKENHI, Grant No.~17H02946; MEXT KAKENHI Grant No.~17H06437 in Innovative Areas ``Innovations for Light--Energy Conversion,'' and MEXT Quantum Leap Flagship Program Grant Nos.~JPMXS0118069242.
\end{acknowledgements}

\appendix

\section{Derivation of the quantum state of entangled three photons via cascaded PDC}
\label{sec:appendix1}
\renewcommand{\theequation}{\ref{sec:appendix1}.\arabic{equation}}
\setcounter{equation}{0}

For simplicity, we consider the electric fields inside the one-dimensional nonlinear crystals. 
The positive-frequency component of the electric field operator is expressed as
\begin{align}
	\E_j(z,t)
	=
	C
	\int d\omega
	e^{i k_j(\omega)z -i\omega t}
	\hat{a}_j(\omega),
\end{align}
where the slowly varying envelope approximation has been adopted, and all constants have been combined into a factor $C$.
We treat the strong electric field of the input laser as a classical field
\begin{align}
	E_\p(z,t)
	=
	\int d\omega
	e^{i k_\p(\omega)z -i\omega t}
	A_\p (\omega),
\end{align}
where $A_\p (\omega)$ denotes the envelope of the input laser.
The Hamiltonian govering the cascaded PDC process can be written as \cite{Zhang:2018ev,Ye:2020dz}
\begin{align}
	\hat{H}_\text{CPDC}(t)
	=
	\hat{H}_1(t)
	+
	\hat{H}_2(t).
\end{align}
In the above equation, $\hat{H}_n(t)$ is the Hamiltonian for the PDC process in the $n$-th crystal
\begin{align}
	\hat{H}_1(t)
	&= 
	\int_{-L_1 /2}^{L_1 /2} dz
	\chi_1^{(2)}
	E_p(z,t)
	\E_0^\dagger(z,t)
	\E_1^\dagger(z,t)
\notag \\
	&\quad
	+
	\text{h.c.},
\end{align}
\begin{align}
	\hat{H}_2(t)
	&= 
	\int_{-L_2 /2}^{L_2 /2} dz
	\chi_2^{(2)}
	\E_0(z,t)
	\E_2^\dagger(z,t)
	\E_3^\dagger(z,t)
\notag \\
	&\quad
	+
	\text{h.c.},
\end{align}
where $\chi_n^{(2)}$ is the second-order susceptibility of the $n$-th crystal.
We perform the integration over the length of the crystal and obtain
\begin{align}
	\hat{H}_1(t)
	&=
	C_1
	\iiint d\omega^3
	\phi_1(\omega,\omega_0,\omega_1)
	e^{-i (\omega - \omega_0 -\omega_1) t}
\notag \\
	&\quad \times
	A_\p (\omega)
	\hat{a}_0^\dagger(\omega_0) 
	\hat{a}_1^\dagger(\omega_1) 
	+
	\text{h.c.},
	\label{eq:Hamiltonian-PDC1}
\end{align}
\begin{align}
	\hat{H}_2(t)
	&=
	C_2
	\iiint d\omega^3
	\phi_2(\omega_0,\omega_2,\omega_3)
	e^{-i (\omega_0 - \omega_2 -\omega_3) t}
\notag \\
	&\quad \times
	\hat{a}_0(\omega_0) 
	\hat{a}_2^\dagger(\omega_2) 
	\hat{a}_3^\dagger(\omega_3)  
	+
	\text{h.c.},
	\label{eq:Hamiltonian-PDC2}
\end{align}
in terms of the phase-matching functions
\begin{align}
	&\phi_1(\omega,\omega_0,\omega_1)
	=
	\sinc
	\frac{ [k_\p(\omega) - k_0(\omega_0) -k_1(\omega_1) ] L_1}{2},
\\[12pt]
	&\phi_2(\omega_0, \omega_2, \omega_3)
	=
	\sinc
	\frac{[k_0 (\omega_0)-k_2(\omega_2)-k_3(\omega_3)]L_2}{2}.
\end{align}
In Eqs.~\eqref{eq:Hamiltonian-PDC1} and \eqref{eq:Hamiltonian-PDC2}, $C_1$ and $C_2$ accumulate all constants.

The unitary evolution of a state vector can be expressed as
\begin{align}
	\lvert \psi_\text{CPDC} \rangle 
	&=
	\mathrm{T}
	\exp
	\left[
	-\frac{i}{\hbar}
	\int_{-\infty}^\infty dt \hat{H}_\text{CPDC}(t)
	\right]
	\lvert \vac \rangle,
	\label{eq:unitary-operator-T}
\end{align}
where the symbol ``$\mathrm{T}$'' denotes the time-ordering operator.
We focus on the low-downconversion regime.
In this situation, the time-ordering can be ignored,\cite{Quesada:2014ef} and thus the vector state in Eq.~\eqref{eq:unitary-operator-T} is approximately described in the Taylor series as
\begin{align}
	\lvert \psi_\text{CPDC} \rangle 
	&\simeq
	\exp
	\left[
	-\frac{i}{\hbar}
	\int_{-\infty}^{\infty} dt \hat{H}_\text{CPDC}(t)
	\right]
	\lvert \vac \rangle
\notag \\
	&=
	\lvert \vac \rangle
	+
	\left(-\frac{i}{\hbar} \right)
	\int_{-\infty}^\infty dt \hat{H}_{1}(t)
	\lvert \vac \rangle
\notag \\
	&\quad+
	\frac{1}{2!}
	\left(-\frac{i}{\hbar} \right)^2
	\int_{-\infty}^\infty dt'
	\int_{-\infty}^{\infty} dt''
	 \hat{H}_{2}(t')
 	 \hat{H}_{1}(t'')
	\lvert \vac \rangle
\notag \\
	&\quad+
	\frac{1}{2!}
	\left(-\frac{i}{\hbar} \right)^2
	\int_{-\infty}^\infty dt'
	\int_{-\infty}^{\infty} dt''
	 \hat{H}_{1}(t')
 	 \hat{H}_{1}(t'')
	\lvert \vac \rangle
\notag \\
	&\quad+
	\cdots.
	\label{eq:state-vector-Taylor}
\end{align}
The second term in Eq.~\eqref{eq:state-vector-Taylor} represents the photon pair creation in the primary PDC, and the fourth term is the process where photon pairs are generated twice in the first crystal.
These terms are ignored because they do not contribute to the two-photon counting signal between photons~1 and 3.
The third term in Eq.~\eqref{eq:state-vector-Taylor} describes the quantum state of the entangled three photons, which can be computed as
\begin{align}
	\lvert \psi_\cpdc \rangle 
	&=
	\frac{1}{2!}
	\left(-\frac{i}{\hbar} \right)^2
	\int_{-\infty}^\infty dt' 
	\int_{-\infty}^\infty dt'' 
	 \hat{H}_{2}(t')
 	 \hat{H}_{1}(t'')
	\lvert \vac \rangle
\notag \\
	&=
	- \frac{C_1 C_2}{2 \hbar^2}
	\int \!\!\! \int \!\!\! \int \!\!\! \int \!\!\! \int d\omega^5
	A_\p (\omega)
	\phi_1(\omega,\omega_0,\omega_1)
	\phi_2(\omega_0,\omega_2,\omega_3)
\notag \\
	&\quad \times
	\int_{-\infty}^\infty dt'
	e^{-i (\omega_0 - \omega_2 -\omega_3) t'}
\notag \\
	&\quad \times
	\int_{-\infty}^\infty dt''
	e^{-i (\omega - \omega_0 -\omega_1) t''}
	\hat{a}_1^\dagger(\omega_1) 
	\hat{a}_2^\dagger(\omega_2) 
	\hat{a}_3^\dagger(\omega_3) 
	\lvert \vac \rangle.
\end{align}
We take the time integral to find $
	\int_{-\infty}^\infty dt''
	e^{-i (\omega - \omega_0 -\omega_1) t''}
	=
	2\pi
	\delta(\omega - \omega_0 -\omega_1)
$ and $
	\int_{-\infty}^\infty dt'
	e^{-i (\omega_0 - \omega_2 -\omega_3) t'}
	=
	2\pi
	\delta(\omega_0 - \omega_2 -\omega_3)
$, which give the energy conservation condition $\omega=\omega_1+\omega_2+\omega_3$.
Finally, the state vector of the generated three photon can be re-expressed as
\begin{align}
	\lvert \psi_\cpdc \rangle 
	&=
	\iiint d\omega^3
	f(\omega_1,\omega_2,\omega_3)
	\hat{a}_1^\dagger(\omega_1) 
	\hat{a}_2^\dagger(\omega_2) 
	\hat{a}_3^\dagger(\omega_3) 
	\lvert \vac \rangle,
\end{align}
where $f(\omega_1,\omega_2,\omega_3)$ is given by Eq.~\eqref{eq:state-vector} in Sec.~IIA.

\begin{figure}
	\includegraphics{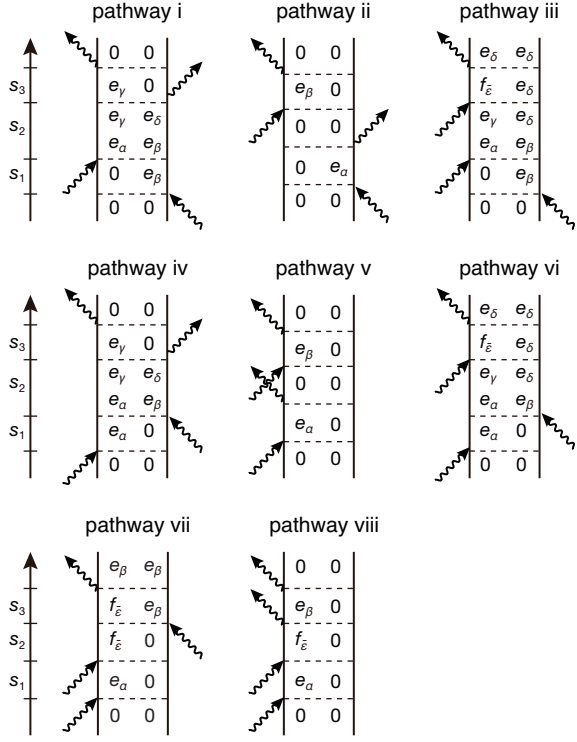}
	\caption{
	Double-sided Feynman diagrams which represent the Liouville space pathways contributing to the third-order responses in the rotating-wave approximation.
	}
	\label{fig:8}
\end{figure}

\section{Derivation of two-photon coincidence counting signal}
\label{sec:appendix3}
\renewcommand{\theequation}{\ref{sec:appendix3}.\arabic{equation}}
\setcounter{equation}{0}

We consider a system comprising molecules and light fields.
The total Hamiltonian is written as
\begin{align}
	\hat{H}_{\rm total}
	=
	\hat{H}_{\rm mol}
	+
	\hat{H}_{\rm field}
	+
	\hat{H}_{\rm mol-field}.
\end{align}
The first term represents the Hamiltonian of photoactive degrees of freedom in the molecules, and the second term is the free Hamiltonian of the field.
The electronic states are grouped into well-separated manifolds: electronic ground states $\lvert 0 \rangle$, the single-excitation manifold $\{ \lvert e_\alpha \rangle \}$, and double-excitation manifold $\{ \lvert f_{\bar\gamma} \rangle \}$. The optical transitions are described by the dipole operator, $\hat{\mu} = \sum_\alpha \mu_{\alpha 0} \lvert 0  \rangle \langle e_\alpha \rvert + \sum_{\alpha \bar\gamma} \mu_{\bar\gamma \alpha} \lvert e_\alpha \rangle \langle f_{\bar\gamma} \rvert$.
Under the rotating-wave approximation, the molecule--field interaction can be written as $\hat{H}_\text{mol--field}(t) = -\hat{\mu}^\dagger \E(t) - \hat{\mu} \E^\dagger(t)$.

The density operator of the total system, $\hat{\rho}(t)$, is expanded in powers of the molecular-field interaction, $\hat{H}_\text{mol--field}(t)$, i.e.,\cite{Dorfman:2016da}
\begin{align}
	\hat{\rho}(t)
	=
	\hat{\rho}^{(1)}(t)
	+
	\hat{\rho}^{(2)}(t)
	+
	\hat{\rho}^{(3)}(t)
	+
	\cdots,
\end{align}
where
\begin{align}
	\hat{\rho}^{(n)}(t)
	&=
	\left(
	-\frac{i}{\hbar}
	\right)^n
	\int_{-\infty}^t d\tau_n
	\int_{-\infty}^{\tau_n} d\tau_{n-1}
	\cdots
	\int_{-\infty}^{\tau_2} d\tau_1
\notag \\
	&\quad \times
	\hat{G}(t-\tau_n)
	\hat{H}_\text{mol--field}^\times(\tau_n)
	\hat{G}(\tau_n-\tau_{n-1})
\notag \\
	&\quad \times
	\hat{H}_\text{mol--field}^\times(\tau_{n-1})
	\cdots
	\hat{G}(\tau_2-\tau_1)
\notag \\
	&\quad \times
	\hat{H}_\text{mol--field}^\times(\tau_1)
	\hat{\rho}(-\infty)
	\label{eq:density-operator-general}
\end{align}
with the initial condition $\hat\rho(-\infty) = \lvert 0 \rangle\langle 0 \rvert \otimes \lvert \psi_\cpdc \rangle\langle \psi_\cpdc \rvert$.
In this equation, $\hat{G}(t)$ denotes the Liouville space time-evolution operator used to describe the molecular excitation and the superoperator notation, $\hat{O}_1^\times \hat{O}_2=[\hat{O}_1,\hat{O}_2]$, has been introduced for any operators $\hat{O}_1$ and $\hat{O}_2$. 
We change the time variables: $s_1= \tau_2 - \tau_1$, $\cdots$, $s_{n-1}= \tau_{n} - \tau_{n-1}$ and $s_n= t - \tau_n$.
Thus, Eq.~\eqref{eq:density-operator-general} can be re-expressed as
\begin{align}
	\hat{\rho}^{(n)}(t)
	&=
	\left(
	-\frac{i}{\hbar}
	\right)^n
	\int_{0}^\infty ds_n
	\int_{0}^{\infty} ds_{n-1}
	\cdots
	\int_{0}^{\infty} ds_1
\notag \\
	&\quad \times
	\hat{G}(s_n)
	\hat{H}_\text{mol--field}^\times(t - s_n)
	\hat{G}(s_{n-1})
\notag \\
	&\quad \times
	\hat{H}_\text{mol--field}^\times(t - s_n - s_{n-1})
	\cdots
	\hat{G}(s_1)
\notag \\
	&\quad \times
	\hat{H}_\text{mol--field}^\times(t - s_n \cdots - s_1)
	\hat{\rho}(-\infty).
	\label{eq:density-operator-Taylor}
\end{align}

Upon the substitution of Eq.~\eqref{eq:density-operator-Taylor} into Eq.~\eqref{eq:transmission}, we obtain the perturbative expansion of the two-photon counting signal in powers of the molecular-field interaction,
\begin{align}
	S(\omega,\omega_{\rm r};\Delta t)
	&=
	S^{(1)}(\omega,\omega_{\rm r};\Delta t)
	+
	S^{(2)}(\omega,\omega_{\rm r};\Delta t)
\notag \\
	&\quad 
	+
	S^{(3)}(\omega,\omega_{\rm r};\Delta t)
	+
	\cdots,
	\label{eq:transmission-Taylor}
\end{align}
where
\begin{align}
	S^{(n)}(\omega,\omega_{\rm r};\Delta t)
	&=
	{\rm Im} \int^\infty_{-\infty} dt \, e^{i\omega t}
\notag \\
	&\quad \times
	\tr[ \hat{a}_3^\dagger(\omega_{\rm r}) \hat{a}_3(\omega_{\rm r}) \E_1^\dagger(\omega) \hat\mu \hat\rho^{(n)}(t) ].
\end{align}
We suppose that the molecular system is isotropic.
An even-order response function, such as a second-order response function, disappears in an isotropic medium due to symmetry.
Hence, the lowest non-vanishing nonlinear signal contribution is the third-order term, $S^{(3)}(\omega,\omega_{\rm r};\Delta t)$.
In addition, the first-order contribution to the total signal, $S^{(1)}(\omega,\omega_{\rm r};\Delta t)$, can be removed by considering the difference spectrum in Eq.~\eqref{eq:difference-spectrum}, as discussed in Appendix~C.
The resultant third-order signal is expressed as the sum of eight contributions.
Each contribution is written as
\begin{align}
	S_{x}(\omega,\omega_{\rm r};\Delta t)
	&=
	\mathrm{Im}
	\int^\infty_{-\infty}dt\,e^{i\omega t}
	\iiint^\infty_0 d^3s\,
	\Phi_x(s_3, s_2, s_1)
\notag \\
	&\quad \times
	\C_x(\omega, \omega_{\rm r}, t; s_3, s_2, s_1).
	\label{eq:transmission-general}
\end{align}
In Eq.~\eqref{eq:transmission-general}, the function $\Phi_x(s_3,s_2, s_1)$ indicates the third-order response function associated with the Liouville pathways which are represented by the double-sided Feynman diagrams in Fig.~\ref{fig:8}.
The six-body correlation function of field operators, $\C_x(\omega, \omega_{\rm r}, t; s_3, s_2, s_1)$, are computed as follows: 
\begin{align}
	\C_\mathrm{i}&(\omega, \omega_{\rm r} , t; s_3, s_2, s_1) 
\notag \\
	&=
	\langle \E^\dagger(t-s_3-s_2-s_1) \E(t-s_3)
\notag \\
	& \quad \times
	\E^\dagger_1(\omega) \hat{a}_3^\dagger(\omega_{\rm r}) \hat{a}_3(\omega_{\rm r}) \E(t-s_3-s_2) \rangle,
	\label{eq:six-body-correlation-function}	
\\[12pt]
	\C_\mathrm{ii}&(\omega, \omega_{\rm r} , t; s_3, s_2, s_1) 
\notag \\
	&=
	\langle \E^\dagger(t-s_3-s_2-s_1) \E(t-s_3-s_2)
\notag \\
	& \quad \times
	\E^\dagger_1(\omega) \hat{a}_3^\dagger(\omega_{\rm r}) \hat{a}_3(\omega_{\rm r}) \E(t-s_3) \rangle,
\\[12pt]
	\C_\mathrm{iii}&(\omega, \omega_{\rm r} , t; s_3, s_2, s_1) 
\notag \\
	&=
	\langle \E^\dagger(t-s_3-s_2-s_1) \E^\dagger_1(\omega) \hat{a}_3^\dagger(\omega_{\rm r}) \hat{a}_3(\omega_{\rm r})
\notag \\
	& \quad \times
	\E(t-s_3) \E(t-s_3-s_2) \rangle,
\\[12pt]
	\C_\mathrm{iv}&(\omega, \omega_{\rm r} , t; s_3, s_2, s_1) 
\notag \\
	&=
	\langle \E^\dagger(t-s_3-s_2) \E^\dagger_1(\omega) \hat{a}_3^\dagger(\omega_{\rm r}) \hat{a}_3(\omega_{\rm r})
\notag \\
	& \quad \times
	\E^\dagger(t-s_3) \E(t-s_3-s_2-s_1) \rangle,
\\[12pt]
	\C_\mathrm{v}&(\omega, \omega_{\rm r} , t; s_3, s_2, s_1) 
\notag \\
	&=
	\langle \E^\dagger_1(\omega) \hat{a}_3^\dagger(\omega_{\rm r}) \hat{a}_3(\omega_{\rm r}) \E(t-s_3)
\notag \\
	& \quad \times
	\E^\dagger(t-s_3-s_2) \E(t-s_3-s_2-s_1) \rangle,
\\[12pt]
	\C_\mathrm{vi}&(\omega, \omega_{\rm r} , t; s_3, s_2, s_1) 
\notag \\
	&=
	\langle \E^\dagger(t-s_3-s_2) \E^\dagger_1(\omega) \hat{a}_3^\dagger(\omega_{\rm r}) \hat{a}_3(\omega_{\rm r})
\notag \\
	& \quad \times
	\E(t-s_3) \E(t-s_3-s_2-s_1) \rangle,
\\[12pt]
	\C_\mathrm{vii}&(\omega, \omega_{\rm r} , t; s_3, s_2, s_1) 
\notag \\
	&=
	\langle \E^\dagger(t-s_3) \E^\dagger_1(\omega) \hat{a}_3^\dagger(\omega_{\rm r}) \hat{a}_3(\omega_{\rm r})
\notag \\
	& \quad \times
	\E(t-s_3-s_2) \E(t-s_3-s_2-s_1) \rangle,
\\[12pt]
	\C_\mathrm{viii}&(\omega, \omega_{\rm r} , t; s_3, s_2, s_1) 
\notag \\
	&=
	\langle \E^\dagger_1(\omega) \hat{a}_3^\dagger(\omega_{\rm r}) \hat{a}_3(\omega_{\rm r}) \E^\dagger(t-s_3)
\notag \\
	& \quad \times
	\E(t-s_3-s_2) \E(t-s_3-s_2-s_1) \rangle,
\end{align}
where the bracket denotes the expectation value in terms of the three photon state, namely, $\langle \dots \rangle = \langle \psi_\cpdc \rvert \dots \lvert \psi_\cpdc \rangle$.

To obtain a concrete expression of the signal, here the memory effect straddling different time intervals in the response function is ignored.\cite{Ishizaki:2012kf} For example, the rephasing contribution of the ESA signal is written as
$
\Phi_\mathrm{iii}(t_3,t_2, t_1)
=
-(i/\hbar)^3 \sum_{\alpha\beta\gamma\delta\bar\epsilon}
\mu_{\delta\bar \epsilon} \mu_{\bar\epsilon \gamma} \mu_{\alpha 0} \mu_{0 \beta}
G_{\bar\epsilon\delta}(t_3) G_{\gamma\delta \gets \alpha\beta}(t_2) G_{0\beta}(t_1)
$, where $G_{\gamma\delta \gets \alpha\beta}(t)$ is the matrix element of the time-evolution operator defined by $\rho_{\gamma\delta}(t) = \sum_{\alpha\beta} G_{\gamma\delta \gets \alpha\beta}(t-s) \rho_{\alpha\beta}(s)$, and $G_{\alpha\beta}(t)$ describes the time evolution of the $\lvert e_\alpha \rangle\langle e_\beta \rvert$ coherence.
The Fourier-Laplace transform of $G_{\alpha\beta}(t)$ is introduced as $G_{\alpha\beta}[\omega] = \int^\infty_0 dt\,e^{i\omega t} G_{\alpha\beta}(t)$.

To calculate the signal, the six-body correlation functions of the field operators need to be computed.
For example, the six-body correlation function in the rephasing ESA signal is computed as
\begin{align}
	\C_\mathrm{iii}&(\omega, \omega_{\rm r} , t; s_3, s_2, s_1) 
\notag \\
	&=
	r (\omega,\omega_{\rm r})
	e^{ -i \omega t + i \omega s_3 - i(\omega_\p - \omega - \omega_{\rm r}) s_1 -i (\omega - \bar\omega_1) \dt}
\notag \\ 
	&\quad \times
    [
        \D_1(\omega_{\rm r},s_2 - \dt)
        e^{ i (\omega - \bar\omega_1) s_2}
\notag \\ 
	&\quad        
        +
        \D_1(\omega_{\rm r},s_2 + \dt)
        e^{ i (\omega +\omega_{\rm r} - \bar\omega_2 - \bar\omega_3) s_2}
    ],
\end{align}
where the expression of $\D_1(\omega,t)$ has been defined by Eq.~\eqref{eq:Dn-setup1}.
Unlike the ESA and DQC pathways (pathways~iii and vi -- viii), which contains only normally-ordered field correlation functions, the SE and GSB pathways (pathways~i, ii, iv, and v) need to be treated separately. For example, by using the commutation relation $[\E_1(t),\E_1^\dagger(t')]=\delta(t-t')$, the six-body correlation function in the rephasing SE signal in Eq.~\eqref{eq:six-body-correlation-function} can be recast as two normally-ordered six-body correlations plus a four-body correlation function:
\begin{align}
	\C_\mathrm{i}&(\omega, \omega_{\rm r} , t; s_3, s_2, s_1) 
\notag \\
	&=
	\langle 
	\hat{a}_3^\dagger(\omega_{\rm r})
	\E_2^\dagger(t-s_3-s_2-s_1+\dt) 
\notag \\
	& \quad \times
	\E^\dagger_1(\omega)  
	\E_1(t-s_3-s_2)
	\E_2(t-s_3+\dt) 
	\hat{a}_3(\omega_{\rm r}) 
	\rangle
\notag \\	
	&\quad+
	\langle 
	\hat{a}_3^\dagger(\omega_{\rm r}) 
	\E_2^\dagger(t-s_3-s_2-s_1+\dt) 
\notag \\
	& \quad \times
	\E^\dagger_1(\omega) 
	\E_1(t-s_3) 
	\E_2(t-s_3-s_2+\dt) 
	\hat{a}_3(\omega_{\rm r}) 
	\rangle
\notag \\	
	&\quad+
	e^{-i\omega (t-s_3)}
	\langle 
	\hat{a}_3^\dagger(\omega_{\rm r}) 
	\E_2^\dagger(t-s_3-s_2-s_1+\dt) 
\notag \\
	& \quad \times
	\E_2(t-s_3-s_2+\dt) 
	\hat{a}_3(\omega_{\rm r}) 
	\rangle.
	\label{eq:correlation-SE}
\end{align}
Using Eq.~\eqref{eq:Dn-setup1}, Eq.~\eqref{eq:correlation-SE} can be computed as follows:
\begin{align}
	\C_\mathrm{i}&(\omega, \omega_{\rm r} , t; s_3, s_2, s_1) 
\notag \\
	&=
	r (\omega,\omega_{\rm r})
	e^{ -i \omega t + i \omega s_3 - i(\omega_\p - \omega - \omega_{\rm r}) s_1 -i (\omega - \bar\omega_1) \dt}
\notag \\ 
	&\quad \times
	[
        \D_1(\omega_{\rm r},s_2 - \dt)
        e^{ i (\omega - \bar\omega_1) s_2}
\notag \\	
	&\quad+
        \D_1(\omega_{\rm r},s_2 + \dt)
        e^{ i (\omega +\omega_{\rm r} - \bar\omega_2 - \bar\omega_3) s_2}
   	]
\notag \\	
	&\quad+ 
	  \D_2(\omega_{\rm r},s_1)
	  	e^{ -i (\omega_\p - \omega_{\rm r} -  \bar\omega_1) s_1 }
		e^{-i\omega (t-s_3)}.
\end{align}
The first and second terms in the above equation correspond to the normally-ordered six-body correlation contribution, and the third term is the four-body correlation contribution, which is independent of $\dt$.

The two-photon coincidence counting signal in Eq.~\eqref{eq:transmission-Taylor} is finally expressed as
\begin{align}
	S(\omega,\omega_{\rm r};\Delta t)
	&=
	S_\SE(\omega, \omega_{\rm r};\dt) 
    	+
    	S_\GSB(\omega, \omega_{\rm r};\dt) 
\notag \\
    	&\quad +  
    	S_\ESA(\omega, \omega_{\rm r};\dt)
\notag \\
    	&\quad + 
    	S_\DQC(\omega, \omega_{\rm r};\dt)
\end{align}
in terms of the SE, GSB, ESA, and DQC contributions,
\begin{align}
    &S_\SE(\omega, \omega_{\rm r};\dt) 
\notag \\  
    &\quad \quad
    =
    S_\mathrm{i}(\omega, \omega_{\rm r};\dt) 
    +
    S_\mathrm{iv}(\omega, \omega_{\rm r};\dt),
	\label{eq:SE-definition}
\\[12pt]
    &S_\GSB(\omega, \omega_{\rm r};\dt) 
\notag \\  
    &\quad \quad   
    =
    S_\mathrm{ii}(\omega, \omega_{\rm r};\dt) 
    +
    S_\mathrm{v}(\omega, \omega_{\rm r};\dt),
\\[12pt]
    &S_\ESA(\omega, \omega_{\rm r};\dt) 
\notag \\  
    &\quad \quad     
    =
    S_\mathrm{iii}(\omega, \omega_{\rm r};\dt) 
    +
    S_\mathrm{vi}(\omega, \omega_{\rm r};\dt),
	\label{eq:ESA-definition}
\\[12pt]
    &S_\DQC(\omega, \omega_{\rm r};\dt) 
\notag \\  
    & \quad \quad      
    =
    S_\mathrm{vii}(\omega, \omega_{\rm r};\dt) 
    +
    S_\mathrm{viii}(\omega, \omega_{\rm r};\dt).
	\label{eq:DQC-definition}
\end{align}
The expressions in Eqs.~\eqref{eq:SE-definition} -- \eqref{eq:DQC-definition} are given by Eqs.~\eqref{eq:SE-general} -- \eqref{eq:DQC-general} in Sec.~IIB.

\section{Linear absorption contribution}
\label{sec:appendix4}
\renewcommand{\theequation}{\ref{sec:appendix4}.\arabic{equation}}
\setcounter{equation}{0}

The first-order term in Eq.~\eqref{eq:transmission-Taylor} is written as
\begin{align}
	S^{(1)} (\omega, \omega_{\rm r})
	&=
	{\rm Im}
	\sum_{\alpha} 
	\mu_{\alpha 0}^2
	\int_{-\infty}^\infty dt \,
	e^{i\omega t}
	\int_0^\infty ds_1
	G_{\alpha 0}(s_1)
\notag \\
	&\quad \times     
	\langle 
	\E^\dagger_1(\omega) \hat{a}_3^\dagger(\omega_{\rm r}) \hat{a}_3(\omega_{\rm r}) \E(t-s_1)
	\rangle.
    	\label{eq:absorption}
\end{align}
The four-body correlation function of the electric field operator in Eq.~\eqref{eq:absorption} is computed as
$
\langle 
\E^\dagger_1(\omega) \hat{a}_3^\dagger(\omega_{\rm r}) \hat{a}_3(\omega_{\rm r}) \E_1 (t-s_1)
\rangle 
=
e^{-i\omega (t-s_1)}
r(\omega,\omega_{\rm r})^2
$. Consequently, Eq.~\eqref{eq:absorption} is obtained as
\begin{align}
	S^{(1)}  (\omega, \omega_{\rm r})
	=
	    r(\omega,\omega_{\rm r})^2
    {\rm Re}
	\sum_{\alpha} 
	\mu_{\alpha 0}^2
    G_{\alpha 0}[\omega].
    	\label{eq:absorption1}
\end{align}
Equation~\eqref{eq:absorption1} is independent of the frequency of the input laser, $\omega_\p$, and the delay time, $\dt$.
Hence, this contribution to the total signal in Eqs.~\eqref{eq:transmission} can be removed by considering the difference spectrum in Eq.~\eqref{eq:difference-spectrum}.
When $T_\en^{(23)}$ is much shorter than the characteristic timescales of the dynamics under investigation, Eq.~\eqref{eq:absorption1} reduces to
\begin{align}
	S^{(1)}  (\omega)
	&=
	{\rm sinc}^2\frac{(\omega -\bar\omega_1)T_\en^{(01)} }{2}
	{\rm sinc}^2\frac{(\omega -\bar\omega_1)T_{02} }{2}
\notag \\
    	&\quad \times
	{\rm Re}
	\sum_{\alpha} 
	\mu_{\alpha 0}^2
	G_{\alpha 0}[\omega].
	\label{eq:absorption-limit}
\end{align}
Equation~\eqref{eq:absorption-limit} is independent of the frequency of the input laser, $\omega_\p$, the reference frequency, $\omega_\text{r}$, and the delay time, $\dt$.
By removing this constant background, the linear absorption process can be separated from the pump-probe-type two-photon process.

\section{$\dt$-independent terms in SE and GSB contributions}
\label{sec:appendix5}
\renewcommand{\theequation}{\ref{sec:appendix5}.\arabic{equation}}
\setcounter{equation}{0}

The $\dt$-independent terms in Eqs.~\eqref{eq:SE-general} and \eqref{eq:GSB-general} are computed as follows:
\begin{align}
	\Delta &S_\SE^\r(\omega, \omega_{\rm r})
\notag \\
	&=
	-
	{\rm Re}
	\sum_{\alpha\beta\gamma\delta} 
	\mu_{\delta 0} 
	\mu_{\gamma 0} 
	\mu_{\beta 0}
	\mu_{\alpha 0}
	G_{\gamma 0}[\omega]
	G_{\gamma\delta \gets \alpha\beta}[0]
\notag \\
    	&\quad \times     
	\int_0^\infty ds_1
	e^{-i(\omega_\p -\omega_{\rm r} - \bar\omega_1) s_1}
	G_{0\beta}(s_1)
	\D_2(\omega_{\rm r},s_1),
	\label{eq:SE-dt0}
\\[14pt]
	\Delta &S_\SE^\nr(\omega, \omega_{\rm r})
\notag \\
	&=
	-
	{\rm Re}
	\sum_{\alpha\beta\gamma\delta} 
	\mu_{\delta 0} 
	\mu_{\gamma 0} 
	\mu_{\beta 0}
	\mu_{\alpha 0}	
	G_{\gamma 0}[\omega]
    	G_{\gamma\delta \gets \alpha\beta}[0]
\notag \\
    	&\quad \times	
    	\int_0^\infty ds_1
    	e^{i(\omega_\p -\omega_{\rm r} - \bar\omega_1 ) s_1}
	G_{\alpha 0}(s_1)
    	\D_2(\omega_{\rm r},s_1),
\\[14pt]
	\Delta &S_\GSB^\r(\omega, \omega_{\rm r})
\notag \\
	&=
	-
	{\rm Re}
	\sum_{\alpha\beta} 
	\mu_{\beta 0}^2
	\mu_{\alpha 0}^2
	G_{\beta 0}[\omega]
    	\int_0^\infty ds_1
	e^{-i\omega s_1}
	G_{0 \alpha}(s_1)
\notag \\
    	&\quad \times
    	\int_0^\infty ds_2
    	e^{i(\omega-\omega_\p+\omega_{\rm r} +\bar\omega_1 ) (s_2+s_1)}
	\D_2(\omega_{\rm r},s_2+s_1),
\\[14pt]
	\Delta &S_\GSB^\nr(\omega, \omega_{\rm r})
\notag \\
	&=
	-
	r(\omega,\omega_{\rm r})^2
	{\rm Re}
	\sum_{\alpha\beta} 
	\mu_{\beta 0}^2
	\mu_{\alpha 0}^2
	G_{\beta 0}[\omega]
	G_{\alpha 0}[\omega].
	\label{eq:GSB-dt0}
\end{align}
The contributions to the total signal in Eq.~\eqref{eq:transmission-sum} can be removed by considering the difference spectrum in Eq.~\eqref{eq:difference-spectrum}.
In the limits of $T_\en^{(01)}\to 0$, $T_\en^{(23)}\to 0$, and $T_{02}\to 0$, we obtain $r(\omega_1,\omega_3) = 1$ and $\D_2(\omega, t) = \delta(t)$. Hence, Eqs.~\eqref{eq:SE-dt0} -- \eqref{eq:GSB-dt0} are simplified as follows: 
\begin{align}
	\Delta S_\SE^{(y)}(\omega)
	&=
	-
	{\rm Re}
	\sum_{\alpha\beta\gamma\delta} 
	\mu_{\delta 0} 
	\mu_{\gamma 0} 
	\mu_{\beta 0}
	\mu_{\alpha 0}
\notag \\
	&\quad \times
	G_{\gamma 0}[\omega]
	G_{\gamma\delta \gets \alpha\beta}[0],
	\label{eq:SE-dt0-limit}
\\[12pt]
	\Delta S_\GSB^{(y)}(\omega) 
	&=
	-
	{\rm Re}
	\sum_{\alpha\beta} 
	\mu_{\beta 0}^2
	\mu_{\alpha 0}^2
	G_{\beta 0}[\omega]
	G_{\alpha 0}^{(y)}[\omega].
	\label{eq:GSB-dt0-limit}
\end{align}

\section{Influence of the spectral shape on the temporal resolution}
\label{sec:appendix6}
\renewcommand{\theequation}{\ref{sec:appendix6}.\arabic{equation}}
\setcounter{equation}{0}

While typical nonlinear crystals yield sinc-shaped phase-matching functions, the phase-matching functions can be shaped using custom-poling methods in quasi-phase-matched crystals. \cite{Branczyk:2011en}
Here, we investigate the influence of the spectral shape on the temporal resolution of the two-photon counting signal.
For demonstration purposes, we model the phase-matching function as a Gaussian distribution by approximating the sinc function with a Gaussian of the same width:
\begin{multline}
	r^{(\mathrm{Gauss})}(\omega_1,\omega_3)
\\
	= 
    e^{- \gamma  (\omega_1 - \bar\omega_1)^2 T_\en^2}
    e^{-\gamma \left[ ( \omega_1 - \bar\omega_1)T_{02} +( \omega_3 - \bar\omega_3 ) T_\en^{(23)} \right]^2}
	\label{eq:Gaussian-PMF}
\end{multline}
with $\gamma=0.04825$.
We consider the case that satisfy $T_\en \equiv T_\en^{(01)} = T_\en^{(23)} = 2T_{02}$, and obtain
\begin{align}
    &\D_1^{(\mathrm{Gauss})}(\omega, t)
\notag \\
    &\quad=
    \int^\infty_{-\infty}
    \frac{d\xi}{2\pi}
    e^{-i\xi t} 
    r^{(\mathrm{Gauss})}(\xi + \bar\omega_1,\omega)
\notag   \\
    &\quad=
    \frac{1}{\sqrt{2 \pi} \sigma }
    \exp  \left[ - \frac{t^2}{2\sigma^2 } \right]
\notag \\
	&\quad \quad \times    
    \exp  \left[ - \frac{2i}{5} (\omega-\bar\omega_3) t -\frac{4 \gamma T_\en^2}{5}   (\omega-\bar\omega_3)^2  \right]
	\label{eq:Dn-Gaussian}
\end{align}
with the standard deviation $\sigma=\sqrt{5 \gamma /2} T_\en$.
In terms of the full width at half maximum (FWHM), for the Gaussian phase-matching function in Eq.~\eqref{eq:Gaussian-PMF}, the spreading of the interval between the arrival times of photons~1 and 2 at the molecular sample is estimated to be $2\sqrt{2 \ln 2}\sigma \simeq 0.8176 T_\en$.
Whereas, in the case of the sinc-shaped phase-matching functions in Eq.~\eqref{eq:monochromatic-PMF}, the interval between the arrival times of photons~1 and 2 becomes blurred within the time window of $\text{FWHM}=T_\en$, as shown in Fig.~\ref{fig:2}.
This implies that the temporal resolution of the signal can be improved when the spectral distribution of the phase-matching function in Eq.~\eqref{eq:monochromatic-PMF} can be shaped into a Gaussian distribution of the same bandwidth.

\bigskip


%

\end{document}